\documentclass[10pt]{article}
\usepackage{makeidx}
\usepackage{amsmath}
\usepackage{amsfonts}
\usepackage{amssymb}
\usepackage{graphicx}
\usepackage{cite}
\usepackage{xcolor}
\usepackage{amsthm}

\makeatletter
\g@addto@macro\th@plain{\thm@headpunct{:}}
\makeatother
\theoremstyle{plain}
\newtheorem*{acknowledgement*}{Acknowledgement}
\input epsf.sty
\newtheorem{theorem}{Theorem}
\newtheorem{acknowledgement}[theorem]{Acknowledgement}

\makeatletter \@addtoreset{equation}{section}
\renewcommand{\theequation}{\thesection.\arabic{equation}}
\input{tcilatex}
\begin{document}

\title{\textbf{Domain Walls in Topological Tri-hinge Matter}}
\author{L. B. Drissi$^{1,2,3,\ast}$ and E. H. Saidi$^{1,2,3}$ \\
{\small 1-LPHE, Modeling \& Simulations, Faculty of Science,}\\
{\small {Mohammed V University in Rabat, MB 1014 RP, Rabat, Morocco}} \\
{\small 2- CPM, Centre of Physics and Mathematics, Faculty of Science,
Mohammed V University in Rabat, Morocco} \\
{\small 3- Hassan II Academy of Science and Technology, Rabat, Morocco}\\
{\small $\ast$ E-mails of corresponding author:
lalla-btissam.drissi@um5.ac.ma, b.drissi@academiesciences.ma}}
\maketitle

\begin{abstract}
Using a link between graph theory and the geometry hosting higher order
topological matter, we fill part of the missing results in the engineering
of domain walls supporting gapless states for systems with three vertical
hinges. The skeleton matrices which house the particle states responsible
for the physical properties are classified by the Euler characteristic into
three sets with topological index\textbf{\ }$\chi =0,1,2.$ A tri-hinge
hamiltonian model invariant under the composite\textbf{\ }$\boldsymbol{M}_{1}%
\boldsymbol{T}$, $\boldsymbol{M}_{2}\boldsymbol{T}$, $\boldsymbol{M}_{3}%
\boldsymbol{T}$ is built. In this framework\textbf{, }$\boldsymbol{T}$%
\textbf{\ }is the time reversing symmetry obeying\textbf{\ }$\boldsymbol{T}%
^{2}=-I$ and the $\boldsymbol{M}_{i}$\textbf{'}s are the generators of the
three reflections of the dihedral\textbf{\ }$\mathbb{D}_{3}$\textbf{\ }%
symmetry of triangle\textbf{.} To capture the tri-hinge states, candidate
materials are suggested, thus opening up a variety of possibilities for
investigating and designing robust materials against disorder and
deformation.

\end{abstract}

\section{Introduction}

The discovery of topological insulators and superconductors renders the
bulk-boundary correspondence a general concept valid for a wider set
topological phases of matter; part of which is classified by the periodic
Altland-Zirnbauer (AZ) table\textbf{\ }\textrm{\cite{A1,A2,A3}}\textbf{. }In
this\textbf{\ }classification, the ten AZ symmetry classes (AZ matter below)
are generated by the combination of three internal discrete symmetries,
namely the time-reversal (generated by the operator $\boldsymbol{T}$ ), the
charge conjugation ($\boldsymbol{P}$), and the chiral symmetry ($\boldsymbol{%
C}$) \textrm{\cite{A4,A5,A6}}. Since this breakthrough, the concept of
nontrivial topological band structures has been extended to materials in
which other symmetries are also used to protect the topological phases%
\textrm{\ of matter}. This includes discrete translation symmetry of the
crystal lattice, and other crystalline symmetries like rotations and mirrors
protecting gapless boundary states \textrm{\cite{A7,A8,A9}}.

\  \  \  \newline
Recently, a new family of topological phase of matter \textrm{\cite%
{00C1,00C2}} extending the conventional $D/(D-1)$\textrm{,} sometimes termed
as $D/(D-d)$ correspondence with codimension $d$ bigger than $1$, is
characterized by gapless states in codimension $d$ boundary stabilized by a
discrete symmetry $\boldsymbol{g}$ beyond the AZ invariance generated by $%
\boldsymbol{T},\boldsymbol{P}$ and $\boldsymbol{C}$. In this higher order
topological phase (HOT below), the conventional AZ matter appears just as
the topological phase associated with codimension one ($d=1$). As such, HOT
phase in 2D matter has topologically protected gapless states at corners of
2D polygonal matter; the particle states on the surface and along the
boundary edges are gapped. For the case of HOT matter in 3D, we have two
possible locations of the topologically protected states; either on hinges
or at corners. Based on this image, attempts have been made so far to
classify HOT phases of matter by extending methods used in rederiving AZ
table; and also by using other approaches, like the boundary- based approach
of \cite{Khalaf-Phys.Rev.B2018},\textbf{\ }bulk perspectives as in \cite{VOP}
and algebraic methods considered in \textbf{\ }\cite%
{Trifunovic-Brouwer-Phys-RevX9(2019)}.

\textrm{\  \ }\  \  \newline
From the theoretical modeling view point, the D- geometry (skeleton matrix)
hosting HOT matter is some how irregular; and remarkably, it can be thought
of in terms of D- dimensional graphs with gapless modes located at the
intersection of faces or edges.\textbf{\ }This is our observation that we
want to develop here to complete missing results in literature; in
particular in systems with an odd number of hinges and corners. But before
commenting this lack of results; notice that the correspondence HOT
matter/graphs can be illustrated on several examples \textrm{\cite%
{1D,2D,4D,5D,7D,8D,9D}; }for instance in 2D matter with square shape, HOT
states live at the four corners where the four edges of the 2D graph
intersect. In this case, the gapless states and the domain walls that host
them can be interpreted in terms of the well known Jackiw-Rossi state 
\textrm{\cite{3D}} living at vertices of the graph. Notice also that the
four corners are related to each other by a cyclic $\mathbb{Z}_{4}$ symmetry
which plays together with TRS $\boldsymbol{T}$ a crucial role in the
hamiltonian modeling of HOT matter. Likewise for 3D matter having a cubic
shape, the gapless states emerging at the 8 corners of the 3D graph lie on
the domain walls separating different topological phases. It's the same
thing again for 3D cylindrical matter with a square cross section and
infinite z direction (no corner), here HOT states live on hinges; and, as we
will show in this study, one can still talk about a 3D graph representation
although here we have to do with infinite faces. In all the examples given
above and also for close cousin geometries, the number of edges and corners
where HOT states may live is a multiple of four ($n=4r$), while for
different shape of matter such as the ones with three edges and three
corners ($n=3$), the situation is obscure; to our knowledge no explicit
hamiltonian model for HOT matter with three hinges has been built so far due
to the difficulty in the engineering of domain walls hosting gapless states.
A careful inspection of poly-edge systems reveals that similar results to
the above mentioned square, cube and cylinder can only be written down for
those regular 2D polygons with $4r$ corners and those regular 3D polyhedrons
with 4r edges and 4r corners; this is partially due to specific properties
of $\boldsymbol{C}_{4r}^{z}$ generators.

\  \  \  \  \newline
In this paper, we contribute to HOT matter in 3D by filling part of missing
results in the engineering of domain walls supporting gapless states for the
family of those matter systems with geometric shapes having three vertical
hinges. Physically, this concerns 3D materials with three and six bulk
atoms- bondings which allow geometries having three hinges\ shapes.
Theoretically\textbf{,} this family of 3D materials is modeled by a
cylindrical matter system with triangular section and boundary surface $%
\mathcal{S}$ made of three intersecting faces $\mathcal{S}_{i}$ with
different unit normals $\mathbf{n}_{i}$; but vanishing sum $\mathbf{n}_{1}+%
\mathbf{n}_{2}+\mathbf{n}_{3}=0$. HOT phases in this family of materials sit
either at corners or along hinges; to describe them, we use our observation
relating HOT matter to graph theory and then to the Euler characteristic
class $\chi $; so the D- geometry of cylindrical matter is thought of in
terms of a 3D graph which turn out to be of three types classified by $\chi
=0,1,2$. In this graph classification, corners fall in the classes with $%
\chi \neq 0$ while cylinders infinite in z-direction belong to the class $%
\chi =0$. This classification which will be studied in present paper for
tri-hinge systems holds also for other shapes of cylindrical matter
including those considered in \textrm{\cite{00C1,00C2}}. Using results on
graphs and Dihedral symmetry of triangle, we study the engineering tri-hinge
gapless states by using domain wall approach; we found that hamiltonians
that solve tri-hinge HOT matter constraints have two kinds of symmetries
namely $\boldsymbol{MT}$ for isosceles triangle and \textbf{\ }$\boldsymbol{M%
}_{1}\boldsymbol{T}$,\textbf{\ }$\boldsymbol{M}_{2}\boldsymbol{T}$, $%
\boldsymbol{M}_{3}\boldsymbol{T}$ for equilateral triangle; the $\boldsymbol{%
M}$ refers to a mirror symmetry; it generates a group $\mathbb{Z}_{2}$.

\  \  \  \  \  \newline
The presentation is as follows: In section II, we start by introducing
tri-hinge matter systems and describe briefly some potential materials that
realise them; their bulk atoms have either three bondings or six ones. Then,
we study the link between tri-hinge systems and 3D graphs; this relationship
is presented through illustrating examples used later on. In section III, we%
\textbf{\ }set the basis of hamiltonian modeling of tri-hinge HOT matter;
and in section IV, we develop a hamiltonian model for the class $\chi =0$
and composite\textbf{\ }$\boldsymbol{M}_{1}\boldsymbol{T}$,\textbf{\ }$%
\boldsymbol{M}_{2}\boldsymbol{T}$, $\boldsymbol{M}_{3}\boldsymbol{T}$
symmetries. In the appendix section, we give useful relations concerning the
subgroups of the dihedral symmetry of equilateral triangle (Appendix A);
then we describe the hexagonal frame used in our calculations; other
relations are also developed there (Appendix B); \textrm{and finally, we
give a discussion on the properties of deformation operators according to} $%
\boldsymbol{MT}$ charges; that is whether they preserve $\boldsymbol{MT}$
symmetry or they break it (Appendix C).

\section{Tri-hinges and graph theory}

In this section, we use results on graph theory to approach geometric and
topological properties of cylindrical matter with three hinges. We first
introduce tri-hinge systems as stacked layers 2D atomic thin materials with
hexagonal and triangular structures. Then, we study useful geometric aspects
of cylindrical systems as imagined by the schematic pictures of the \textbf{%
Figure \ref{ab}}. Next, we turn to study their topological properties; here,
we think of the cylinder with three hinges as a 3D graph hosting the
particle states, and use Euler characteristic $\chi $ to classify the 3D
cylindrical graphs into three topological classes.

\subsection{Tri-hinge systems}

\begin{figure}[tbph]
\begin{center}
\includegraphics[width=12cm]{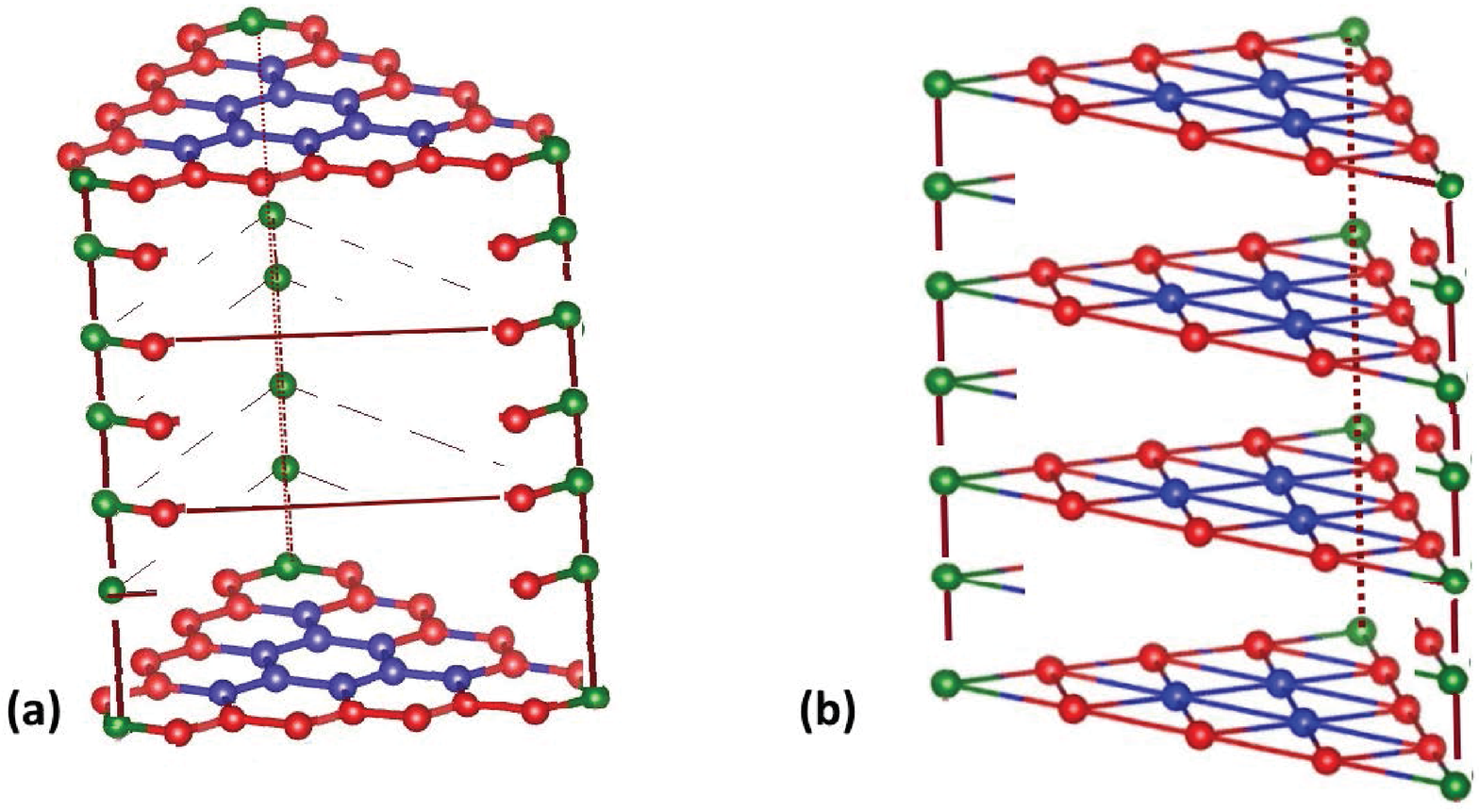}
\end{center}
\par
\vspace{-.5 cm}
\caption{{\protect \small A schematic representation of a cylindrical
material with triangular section. Bulk atoms in blue, surface atoms in red
and hinge atoms in green. This system is made of an AA stacking of several
layers of atoms. Examples of these quasi 2D layers are given by graphene
sheets and counterparts like silicene, boron-nitride and hydrides of atoms
of column IV. They can also be given by materials like Borophene, Alumineme
and homologue having bulk atoms with six bonds.}}
\label{ab}
\end{figure}

Our starting premise is existing or synthetic multilayered structures (MLS)
which result from the reassembly of 2D atomic thin materials through
mechanical stacking in one of the three crystallographic directions. MLS
have been developed as a good alternative to avoid issues limiting the
application of free standing monolayers, such as the strong dependence of
the intrinsic behaviors of these single-layer materials on strain and
substrate. Thus, the stacking of 2D sheets offers an intriguing perspective
of new devices combining exotic phenomena and exploiting the rich physics
involved in their constituents as well as the symmetry of their geometry 
\cite{PRD1,PRD2,fioriA,XiaA,WangA,NPB}\textbf{. }\newline
Among the key parameters for the engineering of the structures composing MLS
is the large possibility in the choice of potential 2D monolayers that are
classified in terms of their elementary bonding states and structures. In
this regard, hexagonal and trigonal materials provide a wide range of basic
building blocks with unique physical properties which do not exist in their
bulk analogs. Graphene is the star member of the honeycomb lattices family
exhibiting ultrathin thickness and characterized by intriguing properties 
\cite{C1,JMP}. Other elements of group-IV, such as silicene, germanene,
stanene, and their hybrids, namely SiC, GeC, SnSi, etc, have also emerged to
widespread the list of 2D hexagonal materials \cite{SI1,GE1,SN1,XC}. In
parallel, the triangular lattices involve group-III 2D materials like
borophene, aluminene and gallenene which have only three valence electrons\ 
\cite{bor1,bor2,AL,ga1,ga2}.

\  \newline
Below we shall think of these MLS as sketched by the two pictures of the 
\textbf{Figure \ref{ab}}; representing stacked layers of 2D materials having
triangular and hexagonal couplings between their in-plane atoms\textbf{. } 
These 3D cylindrical materials with open boundaries in transverse directions
go beyond the family of cubic and rhombohedral materials because of the
difference in the discrete symmetries of their cross sections. Indeed, the
cross sections in Fig(\ref{ab}-a) and (\ref{ab}-b) is triangular or
hexagonal while in cubic and rhombohedral are given by squares or diamond.
Notice that the shape of these sections are known to play an important role
in the engineering of higher order topological states; this is because of
their discrete symmetries of the cross sections; it is these symmetries that
classifies the higher order topological phase of matter \textrm{\cite%
{VOP,nous}}. For example, the studies of topological phases of cylindrical
materials with a square cross section and open boundaries in x- and y-
directions \textrm{\cite{00C1,00C2}} have revealed that the dihedral
symmetry group $\mathbb{D}_{4}$ of the transverse square characterises the
properties of gapless states of higher order; the symmetry z-axis of order 4
that form the $\mathbb{Z}_{4}$ subgroup of $\mathbb{D}_{4}$ characterises
the chiral gapless hinge states; while mirrors generating $\mathbb{Z}_{2}$
subgroups lead to gapless helical states on hinges.

\subsection{Cylindrical matter: geometric aspects}

Generally speaking, 3D cylindrical matter with several hinges are non
regular geometries which can be approached from\textrm{\ }their boundary
properties. The boundary surface of the 3D material is in some sense the
main pillar that encodes the geometric properties of the full 3D system.
Below, we describe rapidly \textrm{four} constituents that play a crucial
role in studying HOT matter and which are related to the boundary surface $%
\mathcal{S}$.

\begin{description}
\item[1)] \  \  \emph{the bulk} $\mathcal{B}$ \emph{with} $\mathcal{S}%
=\partial \mathcal{B}$\newline
The bulk of the matter system with a boundary $\partial \mathcal{B}$ is
parameterised by three variables namely $\mathbf{r}=\left( x,y,z\right) $
with values in a given subset of the 3D space; it hosts gapped bulk states $%
\Psi \left( \mathbf{r}\right) $ with dynamics governed by an hamiltonian H
whose structure will be discussed later on; see Eq(\ref{h}). In the
reciprocal space with variables $\mathbf{k}=\left( k_{x},k_{y},k_{z}\right) $%
, the gapped bulk states are described by wave functions $\Psi _{\mathbf{k}%
}^{\pm }$ with positive and negative frequencies; the $\Psi ^{+}$ for
conduction and the $\Psi ^{-}$ for valence bands. These wave functions are
solutions of the eigenvalue equation $H_{\mathbf{k}}\Psi _{\mathbf{k}}=E_{%
\mathbf{k}}\Psi _{\mathbf{k}}$ with non vanishing gap energy 
\begin{equation}
E_{g}\left( \mathcal{B}\right) >0  \label{a}
\end{equation}%
as required by HOT matter. In our analysis, this bulk $\mathcal{B}$ is given
by the interior of the cylindrical pictures filled by the blue atoms in x-y
plane and along z-axis of the Figures \textbf{\ref{ab}}. For our modeling,
we represent these atomic structures by the cylinders of the Figure\textbf{\ %
\ref{e}}. 
\begin{figure}[tbph]
\begin{center}
\includegraphics[width=12cm]{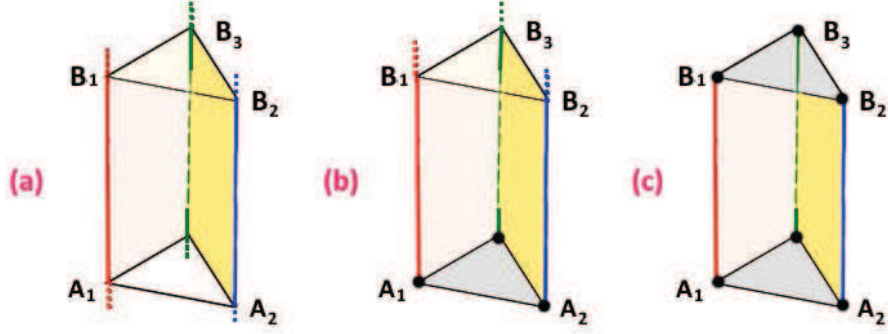}
\end{center}
\par
\vspace{-.5cm}
\caption{{\protect \small Examples of tri-hinge systems with boundary
surfaces }$S${\protect \small \ given a connex union of several faces }$S_{i}$%
{\protect \small : (a) }$S${\protect \small \ is made of three vertical faces
and three hinges; the upper and the lower faces \ are fictitious; they are
pushed to }$\pm \infty ${\protect \small \ or identified by periodicity. (b) }%
$S${\protect \small \ is given by three vertical faces and a horizontal one;
the upper face is pushed to }$+\infty ${\protect \small . Besides hinges,
here we have moreover three vertices represented by black dots. (c) }$S$%
{\protect \small \ is made of five faces; three vertical faces and two
horizontal ones given by the upper and lower full triangles; here we have
six corners}.}
\label{e}
\end{figure}
The graph in the left is infinite in z-direction; it has no corner, the one
in the middle is semi infinite, it has three corners and the graph on the
right side is finite and has 6 corners.

\item[2)] $\  \ $\emph{the boundary surface} $\mathcal{S}$\newline
The boundary surface $\mathcal{S}$ contains the bulk $\mathcal{B}$; for the
pictures of the Figure \textbf{\ref{e},} it is given by a connex union of
several intersecting 2D boundary faces $\mathcal{S}_{i}$. These faces,
thought of below as planar, are parameterised by two independent variables
say $\left( u_{i},v_{i}\right) $; and they play an important role in the
geometrical characterisation of the tri-hinge systems as they can be used to
define all those sub-geometrical objects including hinges and corners. In
the example depicted by the third picture of the Figure \textbf{\ref{e}-c}%
\textrm{\textbf{, }}the connex union of the $\mathcal{S}_{i}$ faces define
the full boundary surface of the material and can be heuristically presented
as follows 
\begin{equation}
\mathcal{S}=\mathcal{S}_{1}+\mathcal{S}_{2}+\mathcal{S}_{3}+\mathcal{S}_{4}+%
\mathcal{S}_{5}
\end{equation}%
This picture represents a 3D system with full open boundaries and so
concerns third order topological phase (TOTIs in language of \textrm{\cite%
{00C223}}) with gapless states at corners. For the Figure \textbf{\ref{e}-c}%
, the boundary surface $\mathcal{S}$ is compact and is made of three
vertical faces $\mathcal{S}_{1},\mathcal{S}_{2},\mathcal{S}_{3}$ and two
horizontal ones $\mathcal{S}_{4},\mathcal{S}_{5}$. An example of a vertical
face is given by the quadrilateral face $\mathcal{F}%
_{A_{1}A_{2}}^{B_{1}B_{2}}$ and a horizontal one by the triangular $\mathcal{%
F}_{A_{1}A_{2}A_{3}}$. \newline
In the cylindrical picture \textbf{\ref{e}-a} with z-axis thought as
infinite, we have three planar surfaces $\mathcal{S}_{1},\mathcal{S}_{2},%
\mathcal{S}_{3}$ with relative orientation angles $\frac{\pi }{3}$; these
surfaces are populated by atoms with red color in the Figure \textbf{\ref{ab}%
}. For HOT matter, surface states have non vanishing gap energy%
\begin{equation}
E_{g}\left( \mathcal{S}\right) >0  \label{b}
\end{equation}

\item[3)] \  \  \emph{the hinges }$\mathfrak{h}_{ij}$\newline
Hinges $\mathfrak{h}_{ij}$ of the cylindrical systems of the Figure \textbf{%
\ref{e}} are defined by the intersection of pairs of faces like 
\begin{equation}
\mathfrak{h}_{ij}=\mathcal{S}_{i}\cap \mathcal{S}_{j}
\end{equation}%
For example, the hinge\ $\mathfrak{h}_{A_{1}A_{2}}$ is given the
intersection of the $\mathcal{F}_{A_{1}A_{2}}^{B_{1}B_{2}}$ and $\mathcal{F}%
_{A_{1}A_{2}A_{3}}$ faces. These $\mathfrak{h}_{ij}$ hinges are
parameterised by one variable say $w_{ij}\sim z$ and are particularly
interesting in the study of the second order topological phase (SOTIs)%
\begin{equation}
E_{g}\left( \mathfrak{h}\right) =0  \label{c}
\end{equation}
as gapless states propagate on these lines. In our situation, we have three
vertical hinges $\mathfrak{h}_{1},\mathfrak{h}_{2},\mathfrak{h}_{3}$\ where
live atoms with a green color as depicted by the Figure \textbf{\ref{ab}. }

\item[4)] \  \  \emph{the corners }$\mathfrak{C}_{ijk}$\newline
Corners $\mathfrak{C}_{ijk}$ in cylindrical systems are as shown by the
pictures (b) and (c) of the Figure \textbf{\ref{e}}; they can be defined in
two ways; either as the intersection of three faces like $\mathfrak{C}_{ijk}=%
\mathcal{S}_{i}\cap \mathcal{S}_{j}\cap \mathcal{S}_{k}$; or as the
intersection of two hinges like $\mathfrak{h}_{ij}\cap \mathfrak{h}_{jk}$.
Corners, where live gapless states and no such states elsewhere, define the
third order topological phase.
\end{description}

\subsection{Graph representation and topology}

Higher topological order phase of 3D systems with three hinges are
particular 3D materials with gapless states living either in codimension
zero (gapless states at corners); or codimension one (gapless states along
hinges). This physical classification brings us to ask a question about the
classification of skeleton matrices inhabiting particle states; this issue
is addressed below.

\subsubsection{Euler characteristic}

\begin{figure}[tbph]
\begin{center}
\includegraphics[width=13cm]{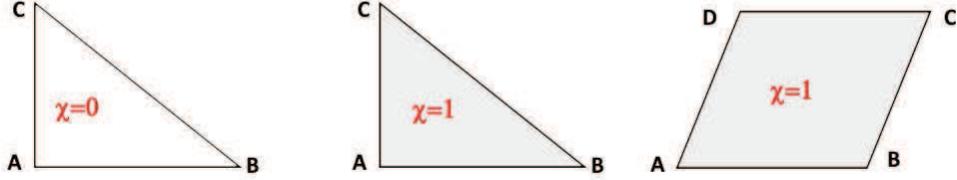}
\end{center}
\par
\vspace{-.5 cm}
\caption{(a) On left an empty triangle with $\protect \chi =0$. (b) In the
middle, a full triangle with $\protect \chi =1$. (c) On right, a
quadrilateral surface with $\protect \chi =1$. The two last surfaces belong
to the same topological class but have different geometries.}
\label{ki}
\end{figure}

From an abstract view, matter skeletons are just particular 3D graphs to
which we can apply known mathematical results such as the Euler
characteristic $\chi $. As such, a classification index of the topological
families of 3D cylindrical matter is given by $\chi \left( \mathcal{S}%
\right) $, the Euler characteristic of the boundary surface\textbf{\ }$%
\mathcal{S}$ of the cylinder with bulk $\mathcal{B}$. Since we deal with a
non conventional boundary surface made of intersecting faces, hinges and
corners; the appropriate index to our calculations is given by the following
integer number\footnote{%
\  \ For any connected planar graph, the Euler characteristic is $F-E+V=2$.
In present study, we work with the relation (\ref{1}) and use it also for
non compact surfaces like for the situation where $F=V=0$.} 
\begin{equation}
\chi \left( \mathcal{S}\right) =F-E+V  \label{1}
\end{equation}%
where $F,$ $E$ and $V$\textrm{\ }refer respectively to the number of faces
making $\mathcal{S}$, the number of its line edges and the number of its
point vertices. For an heuristic illustration of this mathematical formula;
see the three pictures of the Figure \textbf{\ref{ki}}. 
From these pictures, one learns that the empty triangle of the Figure 
\textbf{\ref{ki}}-\textbf{a} belongs to the topological class of the circle $%
\mathbb{S}^{1}$, \textrm{while the Figures} \textbf{\ref{ki}}-\textbf{b} and 
\textbf{\ref{ki}}-\textbf{c}, having $\chi =1,$ belong to the topological
class of a half sphere. \newline
With this tool at hand, we can now turn to our systems. Based on the graph
representation of the shape of matter systems, we can use the Euler
characteristic index (\ref{1}) of graphs to classify the topologies of
cylindric materials. In particular, for\textbf{\ }each $\mathcal{S}_{i}$
face of Figures \textbf{\ref{ki} }corresponds an Euler number equals to
unity; that is $\chi \left( \mathcal{S}_{i}\right) =1$ like the Euler index
of a corner that is also equal to $1$. The situation is different for hinges
that can exist in three types, namely, $\left( i\right) $ infinite hinge
with no ends; its Euler index is equal to $\chi _{h}=-1$; it is negative;
this indicates that the hinge is non compact. $\left( ii\right) $ half
infinite hinge with one end; its Euler index is equal to $\chi _{h}=0$. $%
\left( iii\right) $ finite hinge with two ends; its Euler index is equal to $%
\chi _{h}=+1$, it is positive.

\subsubsection{Topological classes of graphs}

\begin{figure}[tbph]
\begin{center}
\includegraphics[width=10cm]{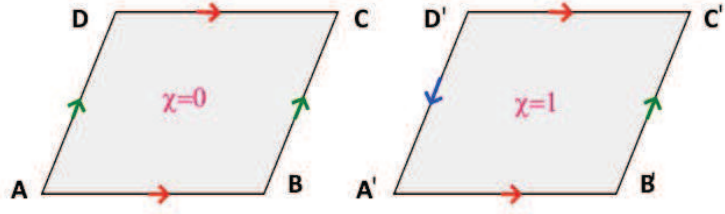}
\end{center}
\par
\vspace{-.5 cm}
\caption{(a) On left the 2-torus where the the parallel edges are identified
by periodicity. (b) On right, only the x- edges are identified by
periodicity; the y-edges have different orientations.}
\label{tor}
\end{figure}

Using eq(\ref{1}), we evaluate the Euler index of the graphs representing
the cylindrical matter of the three pictures of the Figure \textbf{\ref{e}}%
.\ For the Figure \textbf{\ref{e}}-\textbf{c} where the boundary $\mathcal{S}
$ is fully closed, the numbers $\left( F,E,V\right) $ appearing in (\ref{1})
are given by $\left( 5,9,6\right) $, so we have:%
\begin{equation}
\chi \left( \mathcal{S}\right) =2=\chi \left( \mathbb{S}^{2}\right)
\end{equation}%
This value teaches us that the Figure \textbf{\ref{e}-c} is topologically
equivalent to the usual 2-sphere $\mathbb{S}^{2}$ which is known to have a $%
\chi $\ index equal to 2 as given by the index formula $2-2g$ classifying
the Riemann surfaces in terms of the number of genus g (the number of
handles in the surface) \textrm{\cite{RS}}; for the 2-sphere $g=0$ and for
the 2-torus $g=1$. This means that the boundary surface $\mathcal{S}$ of the
Figure \textbf{\ref{e}}-\textbf{c} is topologically equivalent to the
2-sphere; in fact the picture \textbf{\ref{e}-c} is a particular graph
representation of $\mathbb{S}^{2}$ given by three quadrilateral planar
surfaces and two triangular planar ones; other representations are also
possible. In this regards, recall that the geometry of regular 2-sphere can
be\ defined in various but equivalent ways; for example as usual like $%
X^{2}+Y^{2}+Z^{2}=1$ from the view of the 3D space; this expression will be
used below to introduce a cousin surface to $\mathbb{S}^{2}$ namely the real
projective surface $\mathbb{RP}^{2}$. \newline
Regarding the Figure \textbf{\ref{e}-a}, the numbers $\left( F,E,V\right) $
are given by $\left( 3,9,6\right) $; they lead to%
\begin{equation}
\chi \left( \mathcal{S}\right) =0=\chi \left( \mathbb{T}^{2}\right)
\end{equation}%
This configuration is topologically equivalent of the 2- torus $\mathbb{T}%
^{2}$ with genus $g=1$; it corresponds physically to the case of a
parallelogram in x-y plane as the one given by Figur\textbf{e \ref{ki}-c};
but with parallel borders identified by periodic boundary conditions; for
illustration see the Figure\textbf{\  \ref{tor}-a}. 
From this value and the factorisations of $\mathbb{T}^{2}$ and its index $%
\chi \left( \mathbb{T}^{2}\right) $ respectively as the product of two
circles $\mathbb{S}^{1}\times \mathbb{S}^{1}$ and the product of their
characteristics $\chi \left( \mathbb{S}^{1}\right) \chi \left( \mathbb{S}%
^{1}\right) $, it follows that cylinders fall as well in the same
topological class as the 2-torus. Notice that $\chi \left( \mathbb{S}%
^{1}\right) =1-1=0$, takes\textbf{\ }the same value as for the empty
triangle of the Figur\textbf{e \ref{ki}-a}; this can be interpreted as a
deformation of the circle into an empty triangle which is a topological
symmetry of the circle. \newline
For the Figure \textbf{\ref{e}-b}, the numbers $\left( F,E,V\right) $ are
given by $\left( 4,9,6\right) $, so we have 
\begin{equation}
\chi \left( \mathcal{S}\right) =1=\chi \left( \mathbb{RP}^{2}\right)
\end{equation}%
revealing that the boundary surface is topologically equivalent to the real
projective plane $\mathbb{RP}^{2}$. Recall that the surface $\mathbb{RP}^{2}$
is a regular compact surface that can be imagined as close cousin of $%
\mathbb{S}^{2}$; it can be defined by starting for the 2-sphere $%
X^{2}+Y^{2}+Z^{2}=1$ and demanding the identification of the $\left(
X,Y,Z\right) $ points with their antipodes $\left( -X,-Y,-Z\right) $;
roughly, $\mathbb{RP}^{2}$ is just an unoriented half sphere; it sits
between $\mathbb{T}^{2}$ and $\mathbb{S}^{2}$ as shown by the pictures of
the Figure \textbf{\ref{e}}. Notice that as far as topology is concerned,
the value of $\chi \left( \mathbb{RP}^{2}\right) $ is the same as $\chi
\left( D\right) =1$, the Euler characteristics of the disc $D$, and also the
same as the full triangular $\left( ABC\right) $ and the quadrilateral $%
\left( ABCD\right) $ surfaces of the Figure\textbf{\  \ref{ki}}. These
features indicate that although they have different geometric shapes, they
belong to the same topological class as $\mathbb{RP}^{2}$ and so are
topologically equivalent. \newline
Having described the topological aspect of the skeleton matrix hosting
matter interpreted here in terms of particular 3D graphs; we now turn to the
topological properties of the matter itself.

\section{Hamiltonian and domain walls}

So far, we have described topological and geometrical properties of
tri-hinge cylinder made by gluing three vertical planar surfaces with
different orientations. Here, we complete the picture by studying those
topological aspects induced by matter itself; that is by properties of
quantum states moving in bulk, on the boundary surface and especially the
gapless states along hinges of the Figure\textbf{\  \ref{e}-a}.

\subsection{Four bands model}

\begin{figure}[tbph]
\begin{center}
\includegraphics[width=8cm]{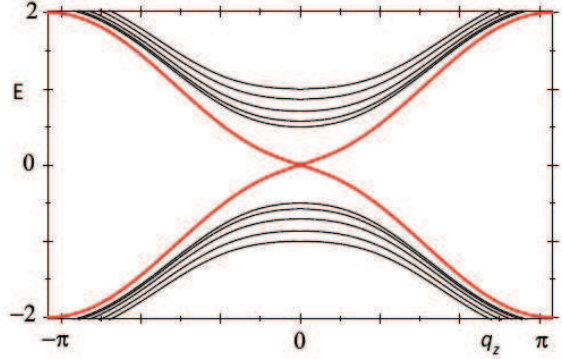}
\end{center}
\caption{{\protect \small Band energies E}$_{\pm }${\protect \small \ as a
function of }$k_{z}${\protect \small \ near the Dirac points } $\left( \pm 
\frac{1}{3},\mp \frac{1}{3},0\right) ${\protect \small \ , the parameters are
as }$\protect \xi =1${\protect \small \ and }$\protect \delta ^{2}=0.25,$%
{\protect \small \ }$0.5,${\protect \small \ }$0.75${\protect \small \ and }$1.$%
}
\label{stb}
\end{figure}

The dynamics of the particle states in 3D cylindrical matter with tri-hinge
is implemented by fixing a quantum model with hamiltonian $H=H_{\mathbf{k}}$
that we comment below and develop it further\textrm{\ }in the next section%
\textbf{.} For concreteness, we take a tight binding model with hermitian 4$%
\times $4 hamiltonian matrix describing 2+2 energy bands: 2 for the valence
band and 2 for the conducting one; see the Figure\textbf{\  \ref{stb} }for
illustration. The general form of this matrix is given by 
\begin{equation}
H=\left( 
\begin{array}{ccc}
H_{11} & \cdots & H_{41} \\ 
\vdots & \ddots & \vdots \\ 
H_{41}^{\dagger } & \cdots & H_{44}%
\end{array}%
\right)  \label{h}
\end{equation}%
it has 16 degrees of freedom $H_{mn}$ which can be approached by expanding
it on a 16-dimensional basis $T^{A}$\ like\textrm{\ }%
\begin{equation}
H_{mn}=F_{0}\delta _{mn}+\sum_{A=1}^{15}F_{A}T_{mn}^{A}  \label{ex}
\end{equation}%
where $F_{A}$'s are functions of momentum\textrm{\ }$\mathbf{k}$ and also of
coupling parameters $\Lambda _{1},...,\Lambda _{r}$ defining the model and
which we denote collectively by $\mathbf{\Lambda }$. The dependence in
momentum is essentially carried by $\sin k_{i}$ and $\cos k_{i}$ functions
coming from Fourier transforms of bulk wave function.\ Practical
achievements of the above expansion is given by the well known 4$\times $4
gamma matrices $\mathbf{\gamma }_{1},\mathbf{\gamma }_{2},\mathbf{\gamma }%
_{3},\mathbf{\gamma }_{4}$ and their products; in particular the quadratic,
the cubic and the quartic products respectively generated by the typical
monomials $\mathbf{\gamma }_{a}\mathbf{\gamma }_{b}$, $\mathbf{\gamma }_{a}%
\mathbf{\gamma }_{b}\mathbf{\gamma }_{c}$ and $\mathbf{\gamma }_{a}\mathbf{%
\gamma }_{b}\mathbf{\gamma }_{c}\mathbf{\gamma }_{d}$ with $a<b<c<d$. 
\textrm{In terms of the gamma matrices }$\mathbf{\gamma }_{a}$\textrm{\ and
their completely antisymmetric product }$\mathbf{\gamma }_{a_{1}}\mathbf{%
\gamma }_{a_{2}}\mathbf{.}..\mathbf{\gamma }_{a_{i}}$\textrm{\ --- that are
conventionally denoted like }$\mathbf{\gamma }_{a_{1}a_{2}...a_{i}}$ \textrm{%
with labels as} $a_{1}<a_{2}<...<a_{i}$---\textrm{, the 16 degrees of
freedom (\ref{ex}) are distributed like 1+4+6+4+1 as illustrated on the
following table}%
\begin{equation}
\begin{tabular}{|l|l|l|l|l|l|}
\hline
$H_{mn}$ & $\boldsymbol{I}$ & $\mathbf{\gamma }_{a}$ & $\mathbf{\gamma }%
_{ab} $ & $\mathbf{\gamma }_{5}\mathbf{\gamma }_{a}$ & $\mathbf{\gamma }_{5}$
\\ \hline
$16$ & $1$ & $4$ & $6$ & $\ 4$ & $1$ \\ \hline
\end{tabular}%
\end{equation}%
\textrm{where }$\boldsymbol{I}$\textrm{\ refers to 4}$\times $\textrm{4
identity matrix and where we have also used the properties }$\mathbf{\gamma }%
_{abc}\sim \mathbf{\gamma }_{d}\mathbf{\gamma }_{5}\varepsilon _{abcd}$%
\textrm{\ and }$\mathbf{\gamma }_{abcd}\sim \varepsilon _{abcd}\mathbf{%
\gamma }_{5}$\textrm{\ with }$\varepsilon _{abcd}$ standing for the
completely antisymmetric Levi-Civita tensor in 4d spaces with non zero value 
$\varepsilon _{1234}=1.$\textrm{\ Regarding these relations, n}otice that
because of the anticommuting property of the gamma matrices \textrm{often
termed as Clifford algebra}%
\begin{equation}
\mathbf{\gamma }_{a}\mathbf{\gamma }_{b}+\mathbf{\gamma }_{b}\mathbf{\gamma }%
_{a}=2\delta _{ab}  \label{cl}
\end{equation}%
there is only one quartic monomial namely $\mathbf{\gamma }_{1}\mathbf{%
\gamma }_{2}\mathbf{\gamma }_{3}\mathbf{\gamma }_{4}$; it is precisely the
so-called ${\small (-)}\mathbf{\gamma }_{5}$ generally used to define
chirality of the particle states; this $\mathbf{\gamma }_{5}$ anticommutes
with the four\textrm{\ }$\mathbf{\gamma }_{a}$'s. Below, we use the
following matrix representation of the gamma matrices 
\begin{equation}
\mathbf{\gamma }_{i}=\tau _{x}\sigma _{i}\quad ,\quad \mathbf{\gamma }%
_{4}=\tau _{y}\sigma _{0}\quad ,\quad \mathbf{\gamma }_{5}=\tau _{z}\sigma
_{0}  \label{ga}
\end{equation}%
it is given by the\textrm{\ tensor} product of two sets of Pauli matrices
namely $\sigma _{i}$ and $\tau _{i}$ with $i=x,y,z$, respectively acting on
spin $\uparrow \downarrow $ and orbital degree of freedom $\mathbf{\phi },$ $%
\mathbf{\chi }$ with $\mathbf{\phi }$ referring to conducting band states
and $\mathbf{\chi }$ \ to the valence ones.\newline
A simple example of the expansion (\ref{ex}) is given by the following
reduced development 
\begin{equation}
H=F_{1}\mathbf{\gamma }_{1}+F_{2}\mathbf{\gamma }_{2}+F_{3}\mathbf{\gamma }%
_{3}+F_{4}\mathbf{\gamma }_{4}+F_{5}\mathbf{\gamma }_{5}  \label{4b}
\end{equation}%
This particular hamiltonian is used here below to derive properties encoded
in the gap energy $E_{g}$. It has only five components $F_{a}=F_{a}\left( 
\mathbf{k},\Lambda \right) ;$ the others have been set to zero for
simplicity; but these zero coefficients might given an interpretation in
terms of symmetry requirements on H. Recall that hamiltonians like (\ref{4b}%
) with less $F$'s or more ones can be classified by using $\boldsymbol{T}$, $%
\boldsymbol{P}$ and $\boldsymbol{C}$ of Altland- Zirnbauer as well as extra
discrete symmetries like mirrors \textrm{\cite{VOP,nous}}. Regarding (\ref%
{4b}), notice that the three first coefficients\textbf{\ }$F_{i}$\textbf{\ }%
\textrm{with}\textbf{\ }$i=x,y,z$ \textrm{are} roughly given by sine
functions like $F_{i}=\Delta _{i}\sin k_{i}$, which\ are odd under $\mathbf{k%
}\rightarrow -\mathbf{k}$, with $\Delta _{i}$ coupling constants interpreted
in terms of energy hopping between closed neighboring sites in the material
lattice; we often refer to these $F_{i}$ three terms as kinetic- like. This
is because in the limit of small wave vector components around $0$ and $\pi $%
, the $\sin k_{i}$ can be replaced by its linear component $F_{i}\sim \pm
k_{i}$; and can be presented as a Dirac- like operator $\vec{F}.\mathbf{\vec{%
\gamma}}$. Regarding the two components $F_{4}$\ and $F_{5}$, they go beyond
the $x,y,z$ directions; they are given by adequate combination of cosine
functions; for the cubic lattice models they can be imagined as $\Delta
_{4/5}+\Delta _{4/5i}^{\prime }\cos k_{i}$; they are even under the change $%
\mathbf{k}\rightarrow -\mathbf{k}$; the $\Delta _{4/5}$ and $\Delta
_{4/5i}^{\prime }$ are extra coupling parameters whose number is fixed by
the symmetry of the model. In the limit of small $k_{i}$, the cosines can be
replaced by $\pm 1$ and so $F_{4}$\ and $F_{5}$ get reduced to constants
that are interpreted in terms of mass- like parameters $m_{4}$ and $m_{5}$
that we comment below; they are functions of the $\Delta _{4/5}$ and $\Delta
_{4/5i}^{\prime }$ coupling parameters.

\subsection{Mass parameters and topological states}

As for the the three $F_{i}=\sin k_{i}$ terms, which make it possible to
locate the Dirac points in the space of momentum, the $m_{4}$ and $m_{5}$
mass like parameters play also an important role in dealing with the
topological properties captured by gapless states. The $m_{5}$ parameter
characterises the nature of the topological phases (trivial or not); and the 
$m_{4}$ allows the engineering\textrm{\footnote{%
\ The sign of m$_{4}$ is important in our construction; it will be used in
section 4 for the implementation of domains walls allowing the engineering
of topological helical hinge states.}} of domain walls; the variation of the
sign change of $m_{4}$ when jumping from a face to its neighboring one
indicates existence of a domain wall on which lives a gapless state.\ To
exhibit how the machinery works, we first calculate the spectrum of the
hamiltonian of the above four band model; in particular its four energy
eigenvalues $E_{1},E_{2},E_{3},E_{4}$. In general this is not a simple task
but for eq(\ref{4b}), they can be straightforwardly obtained due to (\ref{cl}%
); the energy eigenvalues are not all of them different; they have
multiplicity two and are given by $E_{\pm }=\pm \frac{1}{2}E_{g}$ with%
\begin{equation}
E_{g}=2\sqrt{F_{1}^{2}+F_{2}^{2}+F_{3}^{2}+F_{4}^{2}+F_{5}^{2}}  \label{ep}
\end{equation}%
referring to the momentum dependent gap energy. This is a function of
momentum $\mathbf{k}$ and of the coupling parameters $\mathbf{\Lambda }%
\equiv \left( \Lambda _{1},...,\Lambda _{r}\right) $ where r is some
positive integer; these parameters are given here by the $\Delta _{i}$'s,
the $\Delta _{4/5}$ and the $\Delta _{4/5i}^{\prime }$'s. The minimal value
of $E_{g}$ is given by the minima of the five $F_{a}^{2}$ coefficients. In
this regards, the minima of the three first $F_{i}=\sin k_{i}$ terms are
given by $k_{i\ast }=n_{i}\pi $; thus leading to a gap energy 
\begin{equation}
E_{g}^{\left( n_{i}\pi \right) }=2\sqrt{\left. F_{4}^{2}\right \vert
_{n_{i}\pi }+\left. F_{5}^{2}\right \vert _{n_{i}\pi }}
\end{equation}%
which is a function of the coupling parameters $\Lambda _{l}$ of the model;
the $\left. F_{4/5}^{2}\right \vert _{n_{i}\pi }$ refer to $%
F_{4/5}^{2}\left( k_{i\ast }\right) $ with momentum variables $k_{i}$ set to 
$k_{i\ast }$. The topological states near $k_{i\ast }=n_{i}\pi $ correspond
to the situation where $E_{g}^{\left( n_{i}\pi \right) }$, as functions of
the $\Lambda _{l}$'s, vanish at some points $\left. \mathbf{\Lambda }%
^{(p)}\right \vert _{n_{i}\pi }$ in the coupling parameter space. To shed
more light on these gapless states, let us study the case where momentum $%
\mathbf{k}$ is taken around the Dirac point $(n_{x},n_{y},n_{z})=\left(
0,0,0\right) $; that is a hamiltonian of the form%
\begin{equation}
H_{D}=k_{1}\mathbf{\gamma }_{1}+k_{2}\mathbf{\gamma }_{2}+k_{3}\mathbf{%
\gamma }_{3}+m_{4}\mathbf{\gamma }_{4}+m_{5}\mathbf{\gamma }_{5}
\end{equation}%
where we have set $\Delta _{1}=\Delta _{2}=\Delta _{3}=1$. Here, the gap
energy is given by $E_{g}^{0}=E_{+}^{0}-E_{-}^{0}$, which due to (\ref{ep}),
is twice $E_{+}^{0}\left( \mathbf{\Lambda }\right) ;$ so the gapless
condition reads as 
\begin{equation}
E_{g}^{0}\left( \mathbf{\Lambda }\right) =2\sqrt{m_{4}^{2}+m_{5}^{2}}=0
\label{eq}
\end{equation}%
and requires two conditions $m_{4}^{2}=0$ and $m_{5}^{2}=0$. In the next
section, we further develop this study by restricting to tri-hinge systems
and develop a way to realise the constraint relations (\ref{a}) and (\ref{b}%
) as well as the vanishing hinge $E_{g}\left( \mathfrak{h}\right) $
expressed for this Hamiltonian model by the condition $E_{g}\left( \mathbf{%
\Lambda }\right) =0$.

\section{Topological hinge states}

In this section, we use the results obtained above as well as the triangle
symmetries to study the higher topological phase realisations for tri-hinge
systems. For that, we have to specify the codimension of the space where
live the gapless states as we have two possible locations of these massless-
like particles: $\left( i\right) $ at the corners which exist for
topological graphs with $\chi \left( \mathcal{S}\right) \neq 0$; that is for
the case of the cylindrical matter bounded on one z-side or on both z-sides
as given by the Figures \textbf{\ref{e}-b} and\  \textbf{\ref{e}-c}. \textrm{%
This situation is not addressed in this work because we are interested in
the following case: }$\left( ii\right) $\textrm{\ along the hinges which
exist for topological graphs with }$\chi \left( \mathcal{S}\right) =0$%
\textrm{\ as in the case of Figure \ref{e}-a which will be studied in the
following. }

\subsection{Triangular section}

\textrm{In this subsection, we use a group theoretical approach to study the
constraints imposed by second order topological states in a tri-hinge
system. We also give the solution to these constraints for topological
tri-hinges using domain walls ideas; and take the opportunity to comment on
the results obtained in \cite{00C1,00C2,00C223}\ for four-hinge system given
by a cylindrical matter with square cross-section. This extra comment may be
viewed as an alternative approach to rederive the above mentioned results by
using the group theory method.}

\subsubsection{Group theory approach: case tri-hinge and beyond}

Here, we use the group theoretical method to describe symmetry properties of
gapless states that propagate along the tri-hinges of the cylindrical matter
as \textrm{depicted by the Figures \textbf{\ref{ab}} and the Figure} \textbf{%
\ref{e}-a}. To start, notice that, geometrically, the cylinder with a cross
section, given by an equilateral triangular, has z- direction as a symmetry
axis $\boldsymbol{C}_{3}^{z}$ of order 3; \textrm{by threefold order
symmetry we mean that }$\left( \boldsymbol{C}_{3}^{z}\right) ^{3}=I_{id}$%
\textrm{\ and therefore we have the identity }$\left( \boldsymbol{C}%
_{3}^{z}\right) ^{2}=\left( \boldsymbol{C}_{3}^{z}\right) ^{-1}.$ \textrm{In
addition to these rotational symmetries, the cylinder of the Fig \textbf{\ref%
{e}-a} has} \textrm{also three mirror symmetries that we denote like }$%
\boldsymbol{M}_{1},\boldsymbol{M}_{2},\boldsymbol{M}_{3}$; \textrm{these are
three reflections obeying the usual property }$\left( \boldsymbol{M}%
_{i}\right) ^{2}=I_{id}$\textrm{\ and consequently }$\left( \boldsymbol{M}%
_{i}\right) ^{-1}=\boldsymbol{M}_{i}.$ \textrm{So, the cylinder with an
equilateral triangular cross section has a total of six symmetries given by}%
\begin{equation}
\begin{tabular}{llllll}
$I_{id},$ & $C_{3}^{z},$ & $\left( C_{3}^{z}\right) ^{-1},$ & $\boldsymbol{M}%
_{1},$ & $\boldsymbol{M}_{2},$ & $\boldsymbol{M}_{3}$%
\end{tabular}
\label{m3}
\end{equation}%
The symmetry elements \textrm{form} \textrm{precisely the so called} \textrm{%
dihedral symmetry} $\mathbb{D}_{3}$ \textrm{of the equilateral triangle; see
the appendix A for useful details on this finite discrete group}. \textrm{In
this regards, notice that a particularly interesting property of }$\mathbb{D}%
_{3}$\textrm{\ is given by its isomorphism with the symmetric group }$%
\mathbb{S}_{3}$\textrm{\ acting by permuting the components of a set }$%
\Sigma _{3}$\textrm{\ of three elements, say }$\Sigma _{3}=\left \{
1,2,3\right \} $; this isomorphism allows to learn many aspects on the $%
\mathbb{D}_{3}$ of the triangle and its representations without having
recourse to schematic drawings\textrm{. Indeed, the permutation group }$%
\mathbb{S}_{3}$\textrm{\ has 6 operators realised as follows}%
\begin{equation}
\begin{tabular}{llllll}
$I_{id},$ & $\left( 123\right) ,$ & $\left( 132\right) ,$ & $\left(
12\right) ,$ & $\left( 23\right) ,$ & $\left( 31\right) $%
\end{tabular}
\label{31}
\end{equation}%
\textrm{where the three }$\left( 12\right) ,\left( 23\right) $ and $\left(
31\right) $\textrm{\ refer to the transpositions while the two }$\left(
123\right) $\textrm{\ and }$\left( 132\right) $\textrm{\ present the cyclic
permutations. By comparing the two above Eq(\ref{m3}) and Eq(\ref{31}), we
learn that the rotational generator }$C_{3}^{z}$\textrm{\ can be identified
with }$\left( 123\right) $\textrm{, the inverse }$\left( C_{3}^{z}\right)
^{-1}$\textrm{\ with }$\left( 132\right) $\textrm{\ and so on. So the set }$%
\{I_{id},C_{3}^{z},\left( C_{3}^{z}\right) ^{-1}\}$ is nothing but the
subgroup $\mathbb{Z}_{3}$ of the dihedral $\mathbb{D}_{3};$ it is isomorphic
to%
\begin{equation}
\mathbb{Z}_{3}=\left \{ I_{id},\left( 123\right) ,\left( 132\right) \right \}
\label{z3}
\end{equation}%
where $I_{id}$ stands for the identity $\left( 1\right) \left( 2\right)
\left( 3\right) $. \textrm{Before giving other useful symmetry group
theoretical details, notice that topologically speaking}, the total Euler
index $\chi _{_{tot}}$ of the graph associated with the cylinder of the%
\textrm{\ Figure} \textbf{\ref{e}-a} is equal to zero as described before;
this vanishing quantum number can be rederived by computing the total Euler
characteristic $\chi _{tot}$ given by the sum $\sum_{i}\chi \left( \mathcal{S%
}_{i}\right) +\sum_{i}\chi \left( \mathfrak{h}_{i}\right) $; the first block 
$\chi \left( \mathcal{S}_{1}\right) +\chi \left( \mathcal{S}_{2}\right)
+\chi \left( \mathcal{S}_{3}\right) $ corresponds to the contribution of the
three faces which gives $1+1+1=3$; the second block $\sum_{i}\chi \left( 
\mathfrak{h}_{i}\right) $ concerns the contribution of the three infinite
hinges; it is given by $-1-1-1=-3$. So, the total $\chi _{tot}=3-3$ vanishes.

$\bullet $ \emph{domain walls: case tri-hinge }\newline
To engineer gapless states only on the tri-hinges for which the $E_{g}$ gap
energy (\ref{eq}) vanishes on $\mathfrak{h}_{i}$'s and non zero elsewhere,
we use domain walls. \textrm{In this manner of doing (see also footnote 2),
the mass-like term m}$_{4}$\textrm{\ on either side of the hinge }$\mathfrak{%
h}_{i}$\textrm{\ is forced to differ in sign putting constraints on the
symmetry components of }$\mathbb{D}_{4}$\textrm{. So, the }domain walls are
characterised by the sign of the mass parameter \textrm{m}$_{4}\equiv \mu $
whose absolute value is precisely the gap energy; that is $E_{g}=\left \vert
\mu \right \vert $. The mass $\mu $ changes its sign when we cross the
domain wall from left to right and vice versa. As we are looking for gapless
states on the hinges, the domain walls are located at the hinges; then the $%
\mu $ changes its signs when going from a given face $\mathcal{S}_{i}$ to
its neighboring faces $\mathcal{S}_{i-1}$ and $\mathcal{S}_{i+1}$ like%
\begin{equation}
\begin{tabular}{lllllllll}
$\left( 123\right) $ & : & $\mathcal{S}_{1}$ & $\rightarrow $ & $\mathcal{S}%
_{2}$ & $\rightarrow $ & $\mathcal{S}_{3}$ & $\rightarrow $ & $\mathcal{S}%
_{1}$ \\ 
$\left( 132\right) $ & : & $\mathcal{S}_{1}$ & $\rightarrow $ & $\mathcal{S}%
_{3}$ & $\rightarrow $ & $\mathcal{S}_{2}$ & $\rightarrow $ & $\mathcal{S}%
_{1}$%
\end{tabular}
\label{tr}
\end{equation}%
where $C_{3}^{z}=\left( 123\right) $ and $\left( C_{3}^{z}\right)
^{-1}=\left( 132\right) $ are the same as in (\ref{z3}). However, for
tri-hinge matter, we are faced with a remarkable property concerning 3D
systems having an odd number of hinges. As there are three faces, we find
that the domain wall picture cannot be fulfilled by \textrm{sign flip} of
the mass-like parameter $\mu $ \textrm{while demanding symmetry under} $%
\mathbb{Z}_{3}$. Indeed, as we go from a face $\mathcal{S}_{i}$\ with
definite sgn$\left( \mu \right) $; say the face $\mathcal{S}_{1}$ with $%
sgn\left( \mu \right) =+$, and turn around from $\mathcal{S}_{i}$ toward $%
\mathcal{S}_{i+3}\equiv \mathcal{S}_{i}$ by changing $sgn\left( \mu \right) $
whenever we cross from $\mathcal{S}_{j}$ to the surface $\mathcal{S}_{j+1}$,
we end up with a different sign configuration as shown on the following
table 
\begin{equation}
\begin{tabular}{l|lll|lll|}
& $\mathcal{S}_{1}$ & $\mathcal{S}_{2}$ & $\mathcal{S}_{3}$ & $\mathcal{S}%
_{1}$ & $\mathcal{S}_{2}$ & $\mathcal{S}_{3}$ \\ \hline \hline
$sgn\left( \mu \right) $ & $+$ & $-$ & $+$ & $-$ & $+$ & $-$ \\ \hline
$sgn\left( \mu \right) $ & $-$ & $+$ & $-$ & $+$ & $-$ & $+$ \\ \hline
\end{tabular}
\label{sn}
\end{equation}%
For the example of an $\mathcal{S}_{1}$ with a chosen $sgn\left( \mu \right)
=+$ (first row in table (\ref{sn}) ), one can check that that $\mathbb{Z}%
_{3} $ is no longer a symmetry of the domain wall algorithm; indeed after
completing a round tour, the plus sign on $\mathcal{S}_{1}$ gets changed
into $sgn\left( \mu \right) =-$; \textrm{but this behavior contradicts the C}%
$_{3}^{z}$\textrm{\ symmetry which requires} $\left( \mathrm{C}%
_{3}^{z}\right) ^{3}=I_{id}.$ \textrm{This indicates that topological phases
of tri-hinge domain walls cannot be implemented in terms of rotations;
unless we extend the period to two round tours; but this demands higher
order rotation axes like }$C_{6}^{z}$\textrm{\ involving an even number of
faces}. \textrm{In this regards, it is interesting to compare this property
with a known result in the literature regarding the case of the cylinder
considered in \cite{00C1,00C2,00C223} having four hinges and a square cross
section.}

\begin{figure}[tbph]
\begin{center}
\hspace{0cm} \includegraphics[width=14cm]{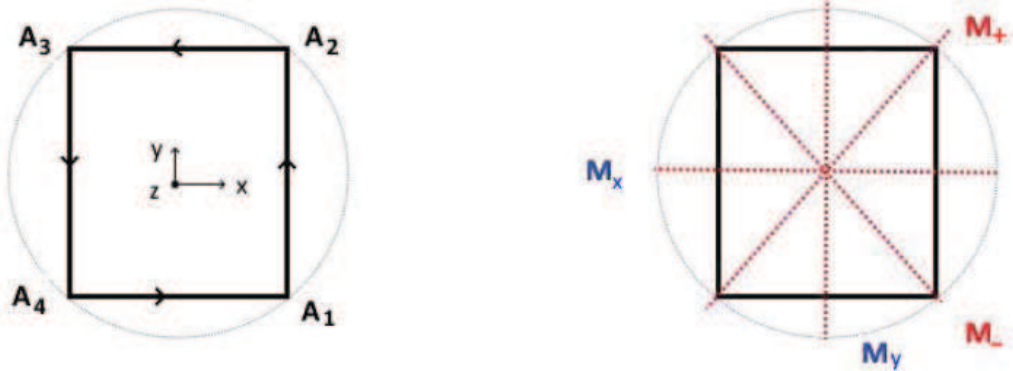}
\end{center}
\par
\vspace{-0.5cm}
\caption{A square cross section with a C$_{4}^{z}$ symmetry. On left the
four corners (hinges) are related by rotations around z-axis with angles $s%
\frac{\protect \pi }{4}$ ---cyclic $\mathbb{Z}_{4}$ symmetry---. On right,
the four mirror symmetries; two of them leave invariant two corners and then
two hinges.}
\label{sq}
\end{figure}

$\bullet $ \emph{domain walls: case four-hinge }\newline
A\textrm{\ schematic representation of cylindric material with square cross
section is depicted by the pictures of the Figure \textbf{\ref{sq}}. } 
\textrm{This open cylindric matter has four hinges }$\mathfrak{h}%
_{1}^{\prime },\mathfrak{h}_{2}^{\prime },\mathfrak{h}_{3}^{\prime },%
\mathfrak{h}_{4}^{\prime }$\textrm{\ and four vertical surfaces }$\mathcal{S}%
_{1}^{\prime },\mathcal{S}_{2}^{\prime },\mathcal{S}_{3}^{\prime },\mathcal{S%
}_{4}^{\prime }$ related among themselves by \textrm{a four order rotation
axis }$C_{4}^{z}$ and its powers $\left( C_{4}^{z}\right) ^{n}$ which obey
the identities $\left( C_{4}^{z}\right) ^{2}=C_{2}^{z}$ and $\left(
C_{4}^{z}\right) ^{3}=\left( C_{4}^{z}\right) ^{-1}$ as it can be checked
from the defining group relation $\left( C_{4}^{z}\right) ^{4}=I_{id}$; the
set of these symmetry operators namely 
\begin{equation}
\mathbb{Z}_{4}=\left \{ I_{id},C_{4}^{z},C_{2}^{z},\left( C_{4}^{z}\right)
^{-1}\right \}  \label{z4}
\end{equation}%
is a symmetry group of the cylinder with square cross section. For this
particular material with square section, \textrm{the homologue of the table (%
\ref{sn}) is given by} 
\begin{equation}
\begin{tabular}{l|llll|llll|}
& $\mathcal{S}_{1}^{\prime }$ & $\mathcal{S}_{2}^{\prime }$ & $\mathcal{S}%
_{3}^{\prime }$ & \multicolumn{1}{l|}{$\mathcal{S}_{4}^{\prime }$} & $%
\mathcal{S}_{1}^{\prime }$ & $\mathcal{S}_{2}^{\prime }$ & $\mathcal{S}%
_{3}^{\prime }$ & $\mathcal{S}_{4}^{\prime }$ \\ \hline \hline
$sgn\left( \mu ^{\prime }\right) $ & $+$ & $-$ & $+$ & \multicolumn{1}{l|}{$%
- $} & $+$ & $-$ & $+$ & $-$ \\ \hline
$sgn\left( \mu ^{\prime }\right) $ & $-$ & $+$ & $-$ & $+$ & $-$ & $+$ & $-$
& $+$ \\ \hline
\end{tabular}
\label{s4}
\end{equation}%
respecting the $\mathbb{Z}_{4}$ symmetry group property $\left(
C_{4}^{z}\right) ^{4}=I_{id}$. Notice also that for the\textrm{\ square
cross section, the symmetry group is larger than the above }$\mathbb{Z}_{4}$%
\textrm{; the full symmetry of the square is given by the dihedral }$\mathbb{%
D}_{4}$ having 8 elements as listed here after\textrm{\ } 
\begin{equation}
\begin{tabular}{llllllll}
$I_{id},$ & $C_{4}^{z},$ & $C_{2}^{z},$ & $\left( C_{4}^{z}\right) ^{-1},$ & 
$\boldsymbol{M}_{x},$ & $\boldsymbol{M}_{y},$ & $\boldsymbol{M}_{+}$ & $%
\boldsymbol{M}_{-}$%
\end{tabular}%
\end{equation}%
This large symmetry group has some remarkable features that make it special
for the study of topological states. Contrary to $\mathbb{D}_{3}$ of the
triangle, the group $\mathbb{D}_{4}$ of the square is not isomorphic to 
\textrm{the symmetric group }$\mathbb{S}_{4}$; but this is not a problem as
there is still an intimate relationship between $\mathbb{S}_{4}$ and $%
\mathbb{D}_{4}$ that can be used to perform suitable calculations; the point
is that $\mathbb{S}_{4}$ has 30 subgroups; three of them are of type $%
\mathbb{D}_{4}$ and nine subgroups of type $\mathbb{Z}_{2}$. In this
regards, recall that $\mathbb{S}_{4}$ has 24 operators which is bigger than
the order 8 of the $\mathbb{D}_{4}$; the group $\mathbb{S}_{4}$ \textrm{acts
by permuting the components of sets }$\Sigma _{4}$\textrm{\ of four
elements, say }$\Sigma _{4}=\left \{ 1,2,3,4\right \} $. To fix ideas, the $%
\Sigma _{4}$ can be imagined in terms of the four corners $%
A_{1},A_{2},A_{3},A_{4}$ of the square as in the \textrm{Figure \textbf{\ref%
{sq}}}; they can be also thought of as the four hinges $\mathfrak{h}%
_{1}^{\prime },\mathfrak{h}_{2}^{\prime },\mathfrak{h}_{3}^{\prime },%
\mathfrak{h}_{4}^{\prime }$ of the cylinder; or the four vertical surfaces $%
\mathcal{S}_{1}^{\prime },\mathcal{S}_{2}^{\prime },\mathcal{S}_{3}^{\prime
},\mathcal{S}_{4}^{\prime }$. Regarding the dihedral $\mathbb{D}_{4}$, it is
a particular subgroup of $\mathbb{S}_{4}$; it has two generators for example
the 4-cycle $\left( 1234\right) $ and the transposition $\left( 13\right) $%
\textrm{. Notice moreover, that the} dihedral $\mathbb{D}_{4}$ itself has in
turn subgroups; it has 10 subgroups collected in the following table%
\begin{equation}
\begin{tabular}{|l|l|l|l|l|l|}
\hline
subgroups & $I_{id}$ & $\mathbb{Z}_{2}$ & $\mathbb{V}_{4}$ & $\mathbb{Z}_{4}$
& $\mathbb{D}_{4}$ \\ \hline
\ number & $1$ & $5$ & $2$ & $1$ & $1$ \\ \hline
\end{tabular}
\label{tb}
\end{equation}%
where $\mathbb{Z}_{p}$ are the usual periodic groups; and where $\mathbb{V}%
_{4}$ \textrm{stands for the so called Klein group of even transpositions.
Notice by the way that the five }$\mathbb{Z}_{2}$\textrm{s are given by the
two sets }$\left \{ I_{id},\boldsymbol{M}_{x}\right \} $ and $\left \{
I_{id},\boldsymbol{M}_{y}\right \} $; the two $\mathbb{Z}_{2}^{\pm }$ given
below by Eq(\ref{m2}) and $\left \{ I_{id},C_{2}^{z}\right \} $\textrm{.
What interest us with the table (\ref{tb}) is that a careful inspection of
the results} of \textrm{\cite{00C1,00C2,00C223}} reveals that it is the
subgroups $\mathbb{Z}_{4}$ of Eq (\ref{z4}) and the two mirror groups 
\begin{equation}
\mathbb{Z}_{2}^{+}=\left \{ I_{id},\boldsymbol{M}_{+}\right \} \qquad
,\qquad \mathbb{Z}_{2}^{-}=\left \{ I_{id},\boldsymbol{M}_{-}\right \}
\label{m2}
\end{equation}%
\textrm{that are behind the engineering of domain walls for cylindrical
matter with a square cross section; they preserve the domain wall
prescription. The }$\mathbb{Z}_{4}$\textrm{\ symmetry encodes data on the
topological chiral states; while the two }$Z_{2}^{\pm }$\textrm{s carry
information the topological helical states}.

$\bullet $ \emph{domain walls: case multi-hinge }\newline
Returning to our tri-hinge system and homologue; and on the light of the
above discussion for tri- and four- hinges, we conclude that in general
cylindric matter with a polygonal cross section --- having an odd number of
hinges, say 2r+1--- cannot accommodate domain walls as exhibited by the
representing table (\ref{sn}) for r=1. Indeed, a regular polygon with 2r+1
corner $\left \{ A_{1},A_{2},...,A_{2r+1}\right \} $ has the dihedral $%
\mathbb{D}_{2r+1}$ as a symmetry group; this discrete symmetry contains
several subgroups among which we have the cyclic $\mathbb{Z}_{2r+1}$,
generated by the 2r+1-\textrm{\ order rotational axis }$C_{2r+1}^{z}$, and
various $\mathbb{Z}_{2}$ mirrors. However, as for the case of the tri-hinge
having a $\mathbb{Z}_{3}$ symmetry, the $\mathbb{Z}_{2r+1}$ transformations
are not compatible with the domain walls algorithm; thus indicating that
there no topological chiral states in such cylindric systems; but may have
topological helical states. In what follows, we explore the case of helical
states in the tri-hinge system; we will show that these topological helical
states are indeed protected by mirror reflections $\boldsymbol{M}_{i}$ of Eq(%
\ref{m3}); but composed with TRS; the mirrors in Eq(\ref{m3}) may be viewed
as the analogue of the ones in Eq(\ref{m2}).

\subsubsection{Mirror symmetries in tri-hinge\  \  \ }

\begin{figure}[tbph]
\begin{center}
\hspace{0cm} \includegraphics[width=14cm]{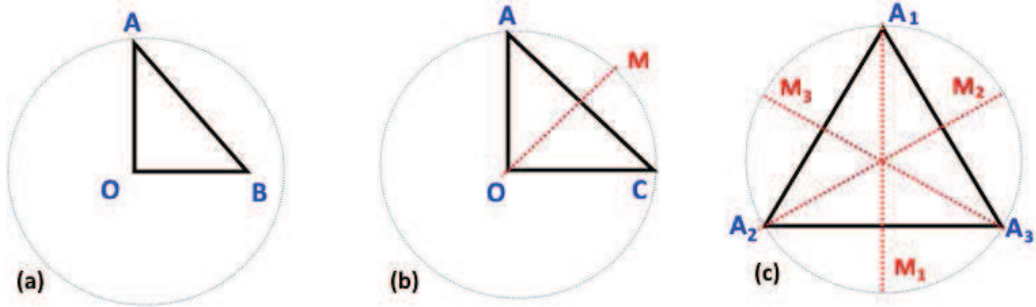}
\end{center}
\par
\vspace{-0.5cm}
\caption{3D system with triangular cross- sections. (a) Triangle with no
reflection symmetry. (b) Isosceles triangle having one reflection symmetry.
(c) equilateral triangle having three reflection symmetries.}
\label{T1}
\end{figure}

We begin by noticing that depending on the geometric shape of the triangular
section, we distinguish three kinds of tri-hinge cylinders. This
classification can be stated in two equivalent manners: $\left( i\right) $
In terms of the values of the three corner angles which have to obey $\alpha
_{1}+\alpha _{2}+\alpha _{3}=\pi $ and solved in three ways as shown by the
second column of the following table,%
\begin{equation}
\begin{tabular}{|l|l|l|l|l|}
\hline
{\small triangle} & {\small corner angles} & \ {\small symmetries} & {\small %
hinge states} & {\small protection} \\ \hline \hline
{\small scalene} & ${\small \alpha }_{1}{\small \neq \alpha }_{2}{\small %
\neq \alpha }_{3}$ & \  \ $\  \  \  \  \ I_{id}$ & {\small gapped} & \  \  \  \ 
{\small -} \\ 
{\small isosceles} & ${\small \alpha }_{1}{\small =\alpha }_{2}{\small \neq
\alpha }_{3}$ & $\  \  \  \ I_{id},\  \boldsymbol{M}$ & {\small gapless} & $\  \ 
\mathbb{Z}_{2}\boldsymbol{T}$ \\ 
{\small equilateral} & ${\small \alpha }_{1}{\small =\alpha }_{2}{\small %
=\alpha }_{3}$ & $I_{id},\  \boldsymbol{M}_{1},\boldsymbol{M}_{2},\boldsymbol{%
M}_{3}$ & {\small gapless} & $\  \  \mathbb{Z}_{2}^{{\small 3}}\boldsymbol{T}$
\\ \hline
\end{tabular}
\label{ta}
\end{equation}%
\ $\left( ii\right) $ In terms of the number of plane symmetries (mirror
reflections $\boldsymbol{M}_{i}$) as given by the third column of the above
table and illustrated by the three pictures of the \textbf{Figure} \textbf{%
\ref{T1}}. 
In this table we have also given the symmetries that protect the gapless
hinge states; these are the $\mathbb{Z}_{2}\boldsymbol{T}$ and $\mathbb{Z}%
_{2}^{3}\boldsymbol{T}$ symmetries of hamiltonian models to be constructed
in subsection 4.2.\newline
The classification of the cross sections of the cylindrical matter can be
also stated by using symmetry group language. Indeed, the largest symmetry
of three objects, say the corners $\left( A_{1},A_{2},A_{3}\right) $ of an
equilateral triangle ---or equivalently the three hinges $\left( \mathfrak{h}%
_{1},\mathfrak{h}_{2},\mathfrak{h}_{3}\right) $; or also the faces $\left( 
\mathcal{S}_{1},\mathcal{S}_{2},\mathcal{S}_{3}\right) $ --- is given by the
dihedral $\mathbb{D}_{3}$ of the equilateral triangle. This group has 6
elements and it is isomorphic to the $\mathbb{S}_{3}$ permutation group
whose elements are given by Eq(\ref{31}). From this group theoretical view,
it is the equilateral triangle which has the full $\mathbb{S}_{3}$ symmetry;
the symmetries of the scalene and the isosceles are subgroups of $\mathbb{D}%
_{3}$; they are as in table (\ref{ta}). For example, the transformation $%
\left( 12\right) $ of the \textbf{Figure} \textbf{\ref{T1}}-b leaves the $%
\mathcal{S}_{3}$ face (edge $CA$) stable and permutes the two other faces $%
\mathcal{S}_{1}$ and $\mathcal{S}_{2}$ (edges $OC$ and $OA$) as follows%
\begin{equation}
\left( 12\right) :\left( \mathcal{S}_{1},\mathcal{S}_{2},\mathcal{S}%
_{3}\right) \rightarrow \left( \mathcal{S}_{2},\mathcal{S}_{1},\mathcal{S}%
_{3}\right)
\end{equation}%
From this algebraic description, we also see that the transformations (\ref%
{tr}) concern the tri-hinge cylinder whose cross section is given by an
equilateral triangle. Using these symmetry properties, it becomes clear why
that the transformations $\left( 123\right) $ and $\left( 132\right) $ are
incompatible with domain walls. The reason is that in the engineering of
domain walls, we need oriented faces $\mathcal{S}_{i}^{\left( +\right) }$
and $\mathcal{S}_{i}^{\left( -\right) }$ with $\left( \pm \right) $
referring to the sign of the mass-like parameter $\mu $. But the
permutations $\left( 123\right) $ and $\left( 132\right) $ do not know about
orientation; they require faces with same orientations; say only $\mathcal{S}%
_{1}^{\left( +\right) },\mathcal{S}_{2}^{\left( +\right) },\mathcal{S}%
_{3}^{\left( +\right) }$; \textrm{this intrinsic structure is consequently
another manifestation of the absence of topological chiral hinge states in
tri-hinge system and generally for an odd number of hinges. Moreover, the
need of both }$\mathcal{S}_{i}^{\left( +\right) }$\textrm{\ and }$\mathcal{S}%
_{i}^{\left( -\right) }$\textrm{\ is in fact a requirement of topological
helical states in tri-hinge; so the engineering of domain walls in tri-hinge
demands the doubling of the degrees of freedom in the original hamiltonian
model. For instance, the modeling by the 4}$\times 4$\textrm{\ matrix of Eqs(%
\ref{h}-\ref{ex}) must be extended to an 8}$\times $\textrm{8; this
construction will be explicitly done in next subsection; it is achieved by
introducing a third set of Pauli matrices denoted as }$\zeta _{x},\zeta
_{y},\zeta _{z}$\textrm{.} \newline
In conclusion, the full $\mathbb{D}_{3}$ symmetry group of equilateral
triangle is incompatible with domain walls' implementation as it breaks the
cyclic $\left( 123\right) $ and its inverse $\left( 132\right) $. Moreover,
knowing that two 3-cycles $\left( 123\right) $ and $\left( 132\right) $
belong to the $\mathbb{Z}_{3}$ subgroup of $\mathbb{D}_{3}$ generated by $%
\boldsymbol{C}_{3}^{z}$, we end up with the result that it is the $\mathbb{Z}%
_{3}$ subset of $\mathbb{D}_{3}$ which is in conflict with the engineering
of domain walls in tri-hinge systems. However, the $\mathbb{D}_{3}\simeq 
\mathbb{S}_{3}$ has other subgroups that may allow domain walls; these are%
\begin{equation}
\begin{tabular}{lllll}
$\mathbb{Z}_{2}$ & : & $I_{id}$ & $;$ & $\left( 12\right) =\boldsymbol{M}%
_{1} $ \\ 
$\mathbb{Z}_{2}^{\prime }$ & : & $I_{id}$ & $;$ & $\left( 23\right) =%
\boldsymbol{M}_{2}$ \\ 
$\mathbb{Z}_{2}^{\prime \prime }$ & : & $I_{id}$ & $;$ & $\left( 31\right) =%
\boldsymbol{M}_{3}$%
\end{tabular}
\label{M3}
\end{equation}%
they correspond to the three mirror reflections depicted by the Figure 
\textbf{\ref{T1}}. For later use we set $\left( ab\right) =\boldsymbol{t}%
_{ab}$; so $\boldsymbol{M}_{1}=\boldsymbol{t}_{12},$ $\boldsymbol{M}_{2}=%
\boldsymbol{t}_{23}$ and $\boldsymbol{M}_{3}=\boldsymbol{t}_{31}$. In what
follows, we show through an explicit construction how these $\mathbb{Z}_{2}$
sub-symmetries of $\mathbb{D}_{3}$ do have indeed domain walls.

\subsection{Domain walls based model}

In this subsection, we consider the tri-hinge cylinder given by the Figure 
\textbf{\ref{e}-a} and construct a domain wall's based model that realises
the second order topological phase of matter. This model is given by the AII
topological class, of the periodic AZ table \cite{A1,A2,A3}, constrained by
the $\mathbb{Z}_{2}$ mirror symmetries of eqs(\ref{M3}). Its hamiltonian $H$
is not invariant under time reversing symmetry (TRS) $\boldsymbol{T}$ alone,
nor under the $\boldsymbol{M}_{i}$ mirrors alone; it is invariant under the
following composed symmetries%
\begin{equation}
\begin{tabular}{lllll}
$\boldsymbol{M}_{1}\boldsymbol{T}$ & , & $\boldsymbol{M}_{2}\boldsymbol{T}$
& , & $\boldsymbol{M}_{3}\boldsymbol{T}$%
\end{tabular}
\label{sy}
\end{equation}%
To that purpose, we first develop the building method; and turn after to
study the implementation of the domain walls and HOT phase in the picture.

\subsubsection{Construction method}

To build the hamiltonian model of the tri-hinge system with second order
topological phase, we start from eq(\ref{4b}) and extend its degrees of
freedom by adding two ingredients: $\left( i\right) $ an extra set of Pauli
matrices $\zeta _{x},\zeta _{y},\zeta _{z}$ together with the usual identity 
$\zeta _{0}$, this is required by the topological helical states; and $%
\left( ii\right) $ a set of projectors $\varrho _{1},\varrho _{2},\varrho
_{3}$ given by $\varrho _{i}=\left \vert i\right \rangle \left \langle
i\right \vert $ with $i=1,2,3$; they are in one to one correspondence with
the three mirror symmetries $\boldsymbol{M}_{1},\boldsymbol{M}_{2},%
\boldsymbol{M}_{3}$. A candidate hamiltonian of the helical tri-hinge model
reads as follows%
\begin{equation}
H=\mathcal{F}_{x}\zeta _{x}\tau _{z}\sigma _{0}+\mathcal{F}_{y}\zeta
_{0}\tau _{x}\sigma _{y}+\Delta _{z}\left( \sin k_{z}\right) \varrho \zeta
_{0}\tau _{x}\sigma _{z}+\mathcal{F}_{4}\zeta _{0}\tau _{y}\sigma _{0}+%
\mathcal{F}_{5}\zeta _{z}\tau _{z}\sigma _{0}  \label{hf}
\end{equation}%
Other candidates for $H$ may be written down; they are given by using
different choices for the block involving the $\zeta _{i}$'s; they will be
briefly commented in appendix C. As other comments \textrm{on} the above
hamiltonian, notice the two following interesting aspects: First, we have
five components $\mathcal{F}_{a}$ to construct; the $\mathcal{F}_{z}$ has
been already replaced by the usual kinetic- like term in reciprocal lattice%
\textrm{\footnote{%
\ The choice of $\sin $ functions is motivated by Dirac theory in small
momentum limit. From the real space view, the tight binding hamiltonian,
relying of the Figure \ref{ab}, involves two kinds of creation $\mathbf{a}_{%
\mathbf{r}}^{\dagger },\mathbf{b}_{\mathbf{r}}^{\dagger }$ and annihilation
operators$\mathbf{a}_{\mathbf{r}},\mathbf{b}_{\mathbf{r}}$ as in the case of
graphene. }} namely $\Delta _{z}\left( \sin k_{z}\right) \varrho $ where $%
\varrho =\varrho _{1}+\varrho _{2}+\varrho _{3}$; this $\mathcal{F}_{z}$ is
odd under $k_{z}\rightarrow -k_{z}$ and commutes with $\boldsymbol{M}$ 
\textrm{as required by TRS and mirror symmetry that we take below as }$\zeta
_{0}\tau _{0}\sigma _{z}$; it vanishes for $k_{z}=n_{z}\pi $ with $n_{z}=0,1$%
. So, it remains four $\mathcal{F}_{a}$'s\ to build. Second, we introduced
three projectors $\varrho _{1},\varrho _{2},\varrho _{3}$; they are needed
to implement the three mirror symmetries (\ref{sy}); for the proof of this;
see eq(\ref{cm}) and the comment following it. The $\mathcal{F}_{a}$'s we
are looking for depend on these projectors; but it turns out that only $%
\mathcal{F}_{y}$ which depends on $\varrho _{i}$ as shown below%
\begin{equation}
\begin{tabular}{lllllll}
$\mathcal{F}_{x}$ & $=$ & $\Delta _{x}F_{x}\varrho $ & \qquad ,\qquad & $%
\mathcal{F}_{4}$ & $=$ & $\Delta _{4}F_{4}\varrho $ \\ 
$\mathcal{F}_{y}$ & $=$ & $\Delta _{y}F_{y}^{i}\varrho _{i}$ & \qquad ,\qquad
& $\mathcal{F}_{5}$ & $=$ & $\Delta _{5}F_{5}\varrho $%
\end{tabular}%
\end{equation}%
where we have also exhibited the hopping energy parameters $\Delta
_{x},\Delta _{y}$ in x- and y- directions as well as coupling constants $%
\Delta _{4},\Delta _{5}$. For simplicity, we set below $\Delta _{x}=\Delta
_{y}=\Delta _{z}=1$. From this parametrisation, we see that we have to
construct $F_{x},F_{y}^{i},F_{4}$ and $F_{5}$ by using the symmetries (\ref%
{sy}). The calculation of these $F_{a}$ is very technical; to organize this
derivation, we split them into two blocks namely: $\left( 1\right) $ the $%
\left( \mathcal{F}_{x},\mathcal{F}_{y}\right) $ block which allows to obtain
the Dirac points in $\left( k_{x},k_{y}\right) $ directions; they will be
studied just below. $\left( 2\right) $ the second block concerns the mass
like components $\left( \mathcal{F}_{4},\mathcal{F}_{5}\right) $; they are
needed to engineer the HOT phase; they will be developed in next sub-
subsection; see \textrm{eqs(\ref{f4})-(\ref{f5})}.

$\bullet $ \emph{the components} $\mathcal{F}_{x},\mathcal{F}_{y}$\newline
To derive the explicit expressions of the components $\mathcal{F}_{x}$ and $%
\mathcal{F}_{y}$, it is interesting to use the 2D hexagonal coordinate
frame; instead of the square coordinates $\left( k_{x},k_{y}\right) $, we
use rather the new momentum variables $\left( q_{1},q_{2},q_{3}\right) $
constrained by the relation 
\begin{equation}
q_{1}+q_{2}+q_{3}=0
\end{equation}%
This parametrisation is interesting as it is invariant under the Dihedral
symmetry group $\mathbb{D}_{3}$ of equilateral triangle. A particular
realisation of these $q_{i}$'s is given by thinking of $q_{3}=-q_{1}-q_{2}$
and $q_{1},q_{2}$ as follows%
\begin{equation}
q_{1}=\frac{k_{x}+\sqrt{3}k_{y}}{2}\quad ,\quad q_{2}=\frac{k_{x}-\sqrt{3}%
k_{y}}{2}  \label{qq}
\end{equation}%
As far as this choice is concerned, we learn from it that $q_{3}=-k_{x}$; we
also learn that the $\left( 12\right) =t_{12}$ transposition, exchanging $%
q_{1}$ and $q_{2},$ corresponds just to $k_{y}\rightarrow -k_{y}$; this
property means that $\mathcal{F}_{y}$ and $\zeta _{0}\tau _{x}\sigma _{y}$
are both of them odd under the mirror $\boldsymbol{M}=\zeta _{0}\tau
_{0}\sigma _{z}$; the oddness of $\mathcal{F}_{y}\left(
q_{1},q_{2},q_{3}\right) $ under $k_{y}\rightarrow -k_{y}$ --- or $\left(
q_{1},q_{2}\right) $ mapped into $\left( q_{2},q_{1}\right) $--- is also a
requirement of TRS represented by $\boldsymbol{T}=\zeta _{0}\tau _{0}\sigma
_{y}\boldsymbol{K}$; see also appendix C for other properties. The oddness
of $\zeta _{0}\tau _{x}\sigma _{y}$ with respect to $\boldsymbol{M}$ is
manifest due to $\sigma _{y}\sigma _{z}=-\sigma _{z}\sigma _{y}$. By taking $%
\mathcal{F}_{x}\left( q_{1},q_{2},q_{3}\right) $ and $\mathcal{F}_{z}\left(
k_{z}\right) $ as well as $\mathcal{F}_{5}\left( q_{1},q_{2},q_{3}\right) $
to be invariant under $\left( q_{1},q_{2}\right) \rightarrow \left(
q_{2},q_{1}\right) $; it follows that $\zeta _{x}\tau _{z}\sigma _{0}$ and $%
\zeta _{0}\tau _{x}\sigma _{z}$ as well as $\zeta _{z}\tau _{z}\sigma _{0}$
should be also invariant under $\boldsymbol{M}=\zeta _{0}\tau _{0}\sigma
_{z} $ which is manifest. This realisation is in agreement with the
classification obtained in \cite{VOP} stating that in general we have two
representations for $\boldsymbol{M}$; one commuting with TRS and the other
one anticommuting with TRS; our $\boldsymbol{M}$ commutes with $\boldsymbol{T%
}$; the other is $\zeta _{y}\tau _{y}\sigma _{y}$; it anticommutes with $%
\boldsymbol{T}$. \newline
\textrm{Regarding} the explicit expressions of $\mathcal{F}_{x},\mathcal{F}%
_{y}$, notice that the constraint $q_{1}+q_{2}+q_{3}=0$ holds everywhere in
our calculations; it can be implemented in the formalism by help of Dirac
delta function $\delta _{q_{1}+q_{2}+q_{3}}$; for simplicity of the
presentation, we shall hide it. Using the new coordinate frame, we can
derive the expressions of the components $\mathcal{F}_{a}$ as functions of
the $q_{i}$'s; the explicit calculations are reported in the appendix B;
they lead to the two following results: $\left( i\right) $ The $\mathcal{F}%
_{x}$ is expressed as $F_{x}\varrho $ with $F_{x}$ given by $\func{Re}Z$
where the complex $Z=e^{2i\pi q_{1}}+e^{2i\pi q_{2}}+e^{2i\pi q_{3}}$ and
its complex adjoint describe transition amplitudes to nearest neighbors. So, 
$F_{x}$ reads as follows%
\begin{equation}
F_{x}=\cos \left( 2\pi q_{1}\right) +\cos \left( 2\pi q_{2}\right) +\cos
\left( 2\pi q_{3}\right)  \label{fx}
\end{equation}%
It is invariant under the dihedral $\mathbb{D}_{3}$ and vanishes for the
values $\left( q_{1\ast },q_{2\ast }\right) =(\frac{1}{3},-\frac{1}{3})$ as
well as $(-\frac{1}{3},\frac{1}{3})$ $\func{mod}1$; these zeros obey $%
q_{1\ast }+q_{2\ast }=0$ $\func{mod}1$. Near, these points, the leading term
in the expansion of $F_{x}$ is given by $\sqrt{3}\left( \tilde{q}_{1}+\tilde{%
q}_{2}\right) /2$ with $\tilde{q}_{i}$ small deviations. $\left( ii\right) $
The component $\mathcal{F}_{y}$ can be presented a the sum $F_{y1}\varrho
_{1}+F_{y2}\varrho _{2}+F_{y3}\varrho _{3}$; it involves a triplet $\left(
F_{y1},F_{y2},F_{y3}\right) $ and its explicit expression is a bit
laborious; for convenience, it is interesting rewrite it like $\varepsilon
_{lij}L_{ij}$ with antisymmetric (triplet) $L_{ij}=L\left(
q_{i},q_{j}\right) $ as follows%
\begin{equation}
\begin{tabular}{lll}
$L_{ij}$ & $=$ & $\cos \left( 2\pi q_{i}\right) -\cos \left( 2\pi
q_{j}\right) $ \\ 
$F_{yl}$ & $=$ & $\varepsilon _{lij}\left[ \cos \left( 2\pi q_{i}\right)
-\cos \left( 2\pi q_{j}\right) \right] $%
\end{tabular}
\label{lij}
\end{equation}%
The tensor $\varepsilon _{lij}$ is the completely antisymmetric Levi-Civita
symbol with $\varepsilon _{123}=1$. Notice the three following features of (%
\ref{lij}): $\left( a\right) $ The $L_{ij}$ is odd under the transpositions $%
\left( ij\right) =\boldsymbol{t}_{ij}$ generating the $\mathbb{Z}_{2}$
mirrors of the $\mathbb{D}_{3}$ symmetry of the triangle; we have: 
\begin{equation}
\boldsymbol{t}_{ij}:L_{ij}\rightarrow L_{ji}=-L_{ij}
\end{equation}%
$\left( b\right) $ The antisymmetric matrix $L_{ij}$ is precisely a triplet $%
\left( L_{12},L_{23},L_{31}\right) $ transforming as a 3-cycle under the $%
\left( 123\right) $ of the $\mathbb{Z}_{3}$ subgroup of the $\mathbb{D}_{3}$%
. This 3-cycle is just the $\mathcal{C}_{3}^{z}$ rotation of the cylinder as
described in appendix A. It acts on the mirror symmetries as%
\begin{equation}
\mathcal{C}_{3}^{z}\boldsymbol{M}_{1}=\boldsymbol{M}_{2}\qquad ,\qquad 
\mathcal{C}_{3}^{z}\boldsymbol{M}_{2}=\boldsymbol{M}_{3}\qquad ,\qquad 
\mathcal{C}_{3}^{z}\boldsymbol{M}_{3}=\boldsymbol{M}_{1}  \label{cm}
\end{equation}%
and teaches us that the realisation of the symmetries eqs(\ref{sy}) is
equivalent to realising $\mathcal{C}_{3}^{z}$ and $\boldsymbol{M}_{1}$.
However, the $\mathcal{C}_{3}^{z}$ is realised by the projectors $\varrho
_{1},\varrho _{2},\varrho _{3}$ as exhibited by eq(\ref{rr}) of appendix B;
it is this property that is behind the use of the $\varrho $'s. $\left(
c\right) $ The set of zeros of the $L_{ij}$ that intersect with those of $%
\mathcal{F}_{1}$ is given by $\left( q_{1\ast },q_{2\ast }\right) =\pm (%
\frac{1}{3},-\frac{1}{3})$; these zeros give the Dirac points where live the
HOT states we are interested in. Near these points, the leading term in the
expansion of $F_{2}$ is given by $\sqrt{3}\left( \tilde{q}_{1}-\tilde{q}%
_{2}\right) /2$.\newline
To exhibit the symmetries (\ref{sy}) of $H$, it is interesting to put it
into a form where the three hinges are apparent. In what follows, we show
that H can be put as follows%
\begin{equation}
H=\left( 
\begin{array}{ccc}
\mathcal{H}\left( q_{1},q_{2}\right) & 0 & 0 \\ 
0 & \mathcal{H}\left( q_{2},q_{3}\right) & 0 \\ 
0 & 0 & \mathcal{H}\left( q_{3},q_{1}\right)%
\end{array}%
\right)
\end{equation}%
where the $\mathcal{H}_{ij}=\mathcal{H}\left( q_{i},q_{j}\right) $ describe
a hinge hamiltonian invariant under $\boldsymbol{M}_{i}\boldsymbol{T}$. As
for $L_{ij}$, the three $\mathcal{H}_{12}$, $\mathcal{H}_{23}$, $\mathcal{H}%
_{31}$ are related to each other by the permutation cycle $\left( 123\right) 
$ generating $\mathbb{Z}_{3}$ subsymmetry of $\mathbb{D}_{3}$.\ To that
purpose, we use the relation $\varrho =\varrho _{1}+\varrho _{2}+\varrho
_{3} $ to decompose the hamiltonian (\ref{hf}) as the sum $H_{1}\varrho
_{1}+H_{2}\varrho _{2}+H_{3}\varrho _{3}$. These $H_{l}$'s differ from each
other only by the component $\mathcal{F}_{y}$; by substituting $%
F_{yl}=\varepsilon _{lij}L_{ij},$ and setting $H_{l}=\varepsilon _{lij}%
\mathcal{H}_{ij}$, we can put the $H_{l}$'s in the equivalent form%
\begin{equation}
\mathcal{H}_{ij}=\Delta _{x}F_{x}\zeta _{0}\tau _{x}\sigma _{x}+\Delta
_{y}L_{ij}\zeta _{0}\tau _{x}\sigma _{y}+\Delta _{z}\left( \sin k_{z}\right)
\varrho \zeta _{0}\tau _{x}\sigma _{z}+\mathcal{F}_{4}\zeta _{0}\tau
_{y}\sigma _{0}+\mathcal{F}_{5}\zeta _{z}\tau _{z}\sigma _{0}
\end{equation}%
where now the $\mathcal{H}_{ij}$ is a function of $\left( q_{i},q_{j}\right) 
$ together with the constraint $q_{1}+q_{2}+q_{3}=0$; i.e $\mathcal{H}_{ij}=%
\mathcal{H}\left( q_{i},q_{j}\right) $ times the Dirac delta function $%
\delta _{q_{1}+q_{2}+q_{3}}$. In this $\mathcal{H}_{ij}$, the components $%
F_{x}$ and $L_{ij}$ are respectively given by (\ref{fx}) and (\ref{lij});
the $\mathcal{F}_{4}$ and $\mathcal{F}_{5}$ are still missing. If we turn
off $\mathcal{F}_{4}$ and $\mathcal{F}_{5}$, the reduced $\mathcal{H}_{ij}|$
is invariant under the time reversing symmetry $\boldsymbol{T}$ represented
by $\zeta _{0}\tau _{0}\sigma _{y}\boldsymbol{K}$ \ and the mirror symmetry $%
\boldsymbol{t}_{ij}$ permuting the variables $q_{i}$ and $q_{j}$; both $%
\boldsymbol{T}$ and $\boldsymbol{t}_{ij}$ leave invariant Dirac delta $%
\delta _{q_{1}+q_{2}+q_{3}}$. However, by turning on $\mathcal{F}_{4}$ and $%
\mathcal{F}_{5}$, the invariance of $\mathcal{H}_{ij}$ is conditioned by the
constraints%
\begin{equation}
\begin{tabular}{lll}
$\mathcal{F}_{4}\left( -q_{i},-q_{j},-k_{z}\right) $ & $=$ & $-\mathcal{F}%
_{4}\left( q_{i},q_{j},k_{z}\right) $ \\ 
$\mathcal{F}_{5}\left( -q_{i},-q_{j},-k_{z}\right) $ & $=$ & $+\mathcal{F}%
_{5}\left( q_{i},q_{j},k_{z}\right) $%
\end{tabular}
\label{c1}
\end{equation}%
required by $\boldsymbol{T}$; and the conditions%
\begin{equation}
\begin{tabular}{lll}
$\mathcal{F}_{4}\left( q_{j},q_{i},k_{z}\right) $ & $=$ & $-\mathcal{F}%
_{4}\left( q_{i},q_{j},k_{z}\right) $ \\ 
$\mathcal{F}_{5}\left( q_{j},q_{i},k_{z}\right) $ & $=$ & $+\mathcal{F}%
_{5}\left( q_{i},q_{j},k_{z}\right) $%
\end{tabular}
\label{c2}
\end{equation}%
demanded by $\boldsymbol{t}_{ij}$. A realisation of these constraints gives
a tri-hinge model invariant under (\ref{sy}). This issue will be studied
after commenting the gap energy since it too is dependent on.

$\bullet $ \emph{gap energy}\newline
The energy eigenvalues of $\mathcal{H}_{ij}$ denoted as $E_{1}^{{\small [ij]}%
},E_{2}^{{\small [ij]}},E_{3}^{{\small [ij]}},E_{4}^{{\small [ij]}}$ are not
all of them different; they have multiplicities and can be presented as $%
E_{\pm }^{{\small [ij]}}=\pm 2E_{g}^{{\small [ij]}}$ with gap energy given by%
\begin{equation}
E_{g}^{{\small [ij]}}=2\sqrt{F_{1}^{2}+L_{ij}^{2}+\sin
^{2}k_{z}+F_{4}^{2}+F_{5}^{2}}  \label{epm}
\end{equation}%
The Dirac points $\left( q_{i\ast },q_{j\ast },k_{z\ast }\right) $ are
obtained by solving $F_{x}^{2}+F_{y}^{2}+\sin ^{2}k_{z}=0$; using previous
results, we end up with the four following points modulo periods 
\begin{equation}
(\frac{1}{3},-\frac{1}{3},0)\quad ,\quad (-\frac{1}{3},\frac{1}{3},0)\quad
,\quad (\frac{1}{3},-\frac{1}{3},\pi )\quad ,\quad (-\frac{1}{3},\frac{1}{3}%
,\pi )
\end{equation}%
Notice that under TRS, the point $(\frac{1}{3},-\frac{1}{3},0)$ is not
invariant; it transforms into $(-\frac{1}{3},\frac{1}{3},0);$ the same
feature holds for $(\frac{1}{3},-\frac{1}{3},\pi )$ which is mapped into $(-%
\frac{1}{3},\frac{1}{3},\pi )$. However, the composition TRS with mirror
symmetry (i.e: $\boldsymbol{MT}$), the Dirac points are invariant as mirror
symmetry permutes $q_{i\ast }$ and $q_{j\ast }$ and leaves $k_{z\ast }$
invariant. At these Dirac points, the above energy eigenvalues reduce down
to $E_{\pm }=\pm 2E_{g}$ with%
\begin{equation}
E_{g}=\frac{1}{2}\sqrt{m_{4}^{2}+m_{5}^{2}}
\end{equation}%
where the mass- like $m_{4},m_{5}$ are the values of $F_{4},F_{5}$ at these
points. To engineer gapless states on the hinges, we need the explicit
expressions of functions $F_{4}$\ and $F_{5}$.

\subsubsection{Implementing domain walls and HOT phase}

Here, we construct mass-like functions $F_{4}$\ and $F_{5}$ solving the
constraint eqs(\ref{c1}-\ref{c2}). In our calculations that follow, we shall
think of $F_{5}$ as the quantity defining the topological phase of the
model; and of the $F_{4}$ as the quantity defining the domain walls needed
by gapless hinge states.\  \newline
We begin by recalling that in the AII topological class of AZ table, time
reversing operator $\boldsymbol{T}$ obeys $\boldsymbol{T}^{2}=-I_{id}$; it
is represented here by the 8$\times $8 matrix operator $\zeta _{0}\tau
_{0}\sigma _{y}\boldsymbol{K}$ where $\boldsymbol{K}$ is the usual complex
conjugation and where $\zeta _{i},\tau _{i}$ and $\sigma _{i}$ are three
sets of Pauli matrices. This $\boldsymbol{T}$ operator, whose inverse $%
\boldsymbol{T}^{-1}$ is equal to $-\boldsymbol{T}$, reads in terms the 4$%
\times $4 gamma matrices used in (\ref{ga}) like $\boldsymbol{T}=i\zeta _{0}%
\mathbf{\gamma }_{1}\mathbf{\gamma }_{3}\boldsymbol{K}$; as such it
anticommutes with $\mathbf{\gamma }_{1},\mathbf{\gamma }_{2},\mathbf{\gamma }%
_{3},\mathbf{\gamma }_{5}$; but commutes with $\mathbf{\gamma }_{4}=\tau
_{y}\sigma _{0}$ due to $\boldsymbol{K\mathbf{\gamma }_{4}}=-\boldsymbol{%
\mathbf{\gamma }_{4}K}$. It is this last property that allows us to
implement domain walls in the construction and then the realisation of a
higher order topological phase.

$\bullet $ \emph{Component }$\mathcal{F}_{4}=\Delta _{4}F_{4}$\newline
Time reversing symmetry requires $F_{4}\left( q_{i},q_{j}\right) $ to be an
odd function under $\boldsymbol{T}$; that is $F_{4}\left(
-q_{i},-q_{j}\right) =-F_{4}\left( q_{i},q_{j}\right) $, a behavior that
cannot be fulfilled as $F_{4}\left( q_{i},q_{j}\right) $ is made of cosine
functions $\cos k_{i}$ which are even under $\boldsymbol{T}$. This
difficulty will be effectively counterbalanced by the use of the geometric
symmetries of HOT phase given by the mirrors (\ref{sy}) as follows%
\begin{equation}
\left( \boldsymbol{MT}\right) \mathcal{H}\left( q_{i},q_{j},k_{z}\right)
\left( \boldsymbol{MT}\right) ^{-1}=\mathcal{H}\left(
-q_{j},-q_{i},-k_{z}\right)  \label{mt}
\end{equation}%
where $\left( \boldsymbol{MT}\right) $ generates a composed TRS-Mirror
symmetry. The reflection operator $\boldsymbol{M}$ acts on momentum $\left(
q_{i},q_{j}\right) $ by permuting the variables and on matrices by the
representation $\zeta _{x}\tau _{y}\sigma _{2}$ \textrm{which reads also as} 
$i\zeta _{x}\mathbf{\gamma }_{y}\mathbf{\gamma }_{5}$; it anticommutes with $%
\zeta _{0}\mathbf{\gamma }_{y}$, commutes with $\zeta _{0}\mathbf{\gamma }%
_{4},\zeta _{z}\mathbf{\gamma }_{5}$ and leaves invariant $\delta
_{q_{1}+q_{2}+q_{3}}$; so we have%
\begin{equation}
\boldsymbol{M}\mathcal{H}\left( q_{i},q_{j},k_{z}\right) \boldsymbol{M}=%
\mathcal{H}\left( q_{j},q_{i},k_{z}\right)
\end{equation}%
An appropriate solution of $F_{4}$ that is independent of momentum $k_{z}$
along the hinges is given by the factorised expression $f_{41}.f_{42}.f_{43}$
with $f_{4i}$ factors given by 
\begin{equation}
\begin{tabular}{lll}
$f_{41}$ & $=$ & $\cos 2\pi q_{1}-\cos 2\pi q_{2}$ \\ 
$f_{42}$ & $=$ & $\cos 2\pi q_{2}-\cos 2\pi q_{3}$ \\ 
$f_{43}$ & $=$ & $\cos 2\pi q_{1}-\cos 2\pi q_{3}$%
\end{tabular}
\label{f4}
\end{equation}%
and $q_{1}+q_{2}+q_{3}=0.$ Notice that under $\boldsymbol{t}_{12}$, the
factor $f_{41}$ changes its sign while $f_{42}$ and $f_{43}$ get
interchanged; the $f_{41}$ vanishes on the hinge $\mathfrak{h}_{12}$ in
agreement with the domain wall prescription. The same feature holds for the
transformations under $\boldsymbol{t}_{23}$ and $\boldsymbol{t}_{31}$. By
substituting $q_{3}=-q_{1}-q_{2}$ and setting $Q_{i}=2\pi q_{i}$, we have%
\begin{equation}
F_{4}=\left( \cos Q_{1}-\cos Q_{2}\right) \left[ \cos Q_{2}-\cos \left(
Q_{1}+Q_{2}\right) \right] \left[ \cos Q_{1}-\cos \left( Q_{1}+Q_{2}\right) %
\right]
\end{equation}%
whose leading term around the Dirac points is $9\sqrt{3}\left( \tilde{q}_{1}-%
\tilde{q}_{2}\right) /8$.

$\bullet $ \emph{Component }$\mathcal{F}_{5}$\newline
Regarding the construction of term $\mathcal{F}_{5}\left(
q_{i},q_{j},k_{z}\right) $, it is even under time reversing symmetry $%
\boldsymbol{T}$ and even under the $\boldsymbol{t}_{ij}$'s. A candidate is
given by%
\begin{equation}
\mathcal{F}_{5}=\Delta _{5}-\frac{\Delta _{5y}^{\prime }}{2\pi }\left[ \cos
2\pi q_{1}+\cos 2\pi q_{2}+\cos 2\pi q_{3}\right] -\Delta _{5z}^{\prime
}\cos k_{z}  \label{f5}
\end{equation}%
where $\Delta _{5},\Delta _{5y}^{\prime }$\textbf{\ }are couplings
constants. The structure of this function resembles to the one of a cylinder
with a square cross section. Nevertheless, the above F$_{5}$ has a special
feature. The particularity of this choice of $F_{5}$ is that at the Dirac
points $\left( q_{1\ast },q_{2\ast }\right) $ given by $\pm \left( \frac{1}{3%
},-\frac{1}{3}\right) $, it reduces to $\xi -\cos k_{z}$ where we have set $%
\xi =\Delta _{5}/\Delta _{5z}^{\prime }$; but this quantity has no
dependence into the coupling $\Delta _{5y}^{\prime }$ at all. This may be
explained as due to the fact that $F_{5}$ can be also presented like 
\begin{equation}
\mathcal{F}_{5}=\Delta _{5}+\Delta _{5y}^{\prime }\frac{\partial F_{y}}{%
\partial q_{2}}-\Delta _{5z}^{\prime }\cos k_{z}
\end{equation}%
with $\frac{\partial F_{y}}{\partial q_{2}}=-2\pi F_{x}$ which vanishes
identically for $\left( q_{1},q_{2}\right) $ equals to $\pm \left( \frac{1}{3%
},-\frac{1}{3}\right) $. Notice also that the reduced expression $\xi -\cos
k_{z}$ has a zero only if $-1\leq \xi \leq 1$; otherwise there is no gapless
state and then no topological insulating phase. Notice moreover that near
the Dirac points $\mathbf{k}_{\ast }$ given by $\left( \pm \frac{1}{3},\mp 
\frac{1}{3},n_{z}\pi \right) $, we can set momentum $\mathbf{k}\sim \mathbf{k%
}_{\ast }+\delta \mathbf{k}$, and approximate the energy eigenvalues (\ref%
{epm}) in nearby of hinges by the quantities $\pm (\delta ^{2}+\sin
^{2}k_{z}+\left( \xi -\cos k_{z}\right) ^{2})^{1/2}$. The variation of these
reduced energies as a function $k_{z}$ is depicted in the Figure \textbf{\ref%
{stb}} for some given values of the parameters $\xi $ and $\delta ^{2}$. In
this figure, we see the gapless mode (in red) traversing the bulk band gap. 

\section{Conclusion and comments}

Higher order topological phase of 2D and 3D matter using domain walls for
the engineering of gapless states in codimension zero and one spaces have
been extensively studied for materials; especially those having cubic and
rhombohedral structures. In this paper\textbf{,} we have contributed to this
matter by completing a missing part of the picture by studying those higher
order topological phases concerning the family of materials having
triangular and hexagonal structures. For this particular tri-hinge systems,
the implementation of domain walls has been a major problem towards the
theoretical construction of explicit Hamiltonian models describing the
higher topological phase. Using the power of the Dihedral symmetry group $%
\mathbb{D}_{3}$ of the triangle and its representations; we have looked for
a group theoretical explanation of the difficulty in the insertion of domain
walls to protect gapless hinge states in a cylindrical system with cross
section given by an equilateral triangle. As a result we have been able to
identify two parts in the Dihedral symmetry $\mathbb{D}_{3}$ of the
triangles. In particular, we found, amongst others, that the $\mathbb{Z}_{3}$
subgroup of $\mathbb{D}_{3}$ forbids \textrm{domain} walls protecting the
chiral hinge states, whilst its three subgroups $\mathbb{Z}_{2\left(
1\right) },$ $\mathbb{Z}_{2\left( 2\right) },$ $\mathbb{Z}_{2\left( 3\right)
}$ allow those protecting helical states. Guided by these symmetry
properties and using the hexagonal frame to deal with the propagation of
quantum states, we have constructed an $\boldsymbol{M}_{i}\boldsymbol{T}$
invariant hamiltonian with manifestly symmetric domain walls. Regarding
applications, we have suggested 3D systems where such HOT phase can be
observed; these materials are given by a cylindrical system with triangular
sections made of the stacking, along the z-axis, of layers of 2D-materials
with triangular and hexagonal structures. In the resulting stacking sheets,
the electronic properties of bulk atoms, forming three and six bonds,
resemble more or less those in their hexagonal and triangular multilayered
analogs respectively. However, the hinge atoms, with only two bonds, leave
an electron quasi free to hop along the hinges direction. \newline
An other interesting finding of this work is the step we have made towards a
classification of higher order topological phases in terms of Euler
characteristics index $\chi $. Though this classification does not
distinguish cylindrical systems with square and triangular sections, it
allows however to foresee three types of cylinders with Euler
characteristics 0,1 and 2 depending on the z-limit of the hinges. Similar
pictures can be drawn for systems with square cross section.

\section{Appendices}

We give three appendices A, B and C: Appendix A collects useful relations
concerning the $\mathbb{D}_{3}$ symmetry of the equilateral triangle.
Appendix B describes the hexagonal coordinates system with applications; and
Appendix C discusses deformations $\delta H_{k}$ preserving\emph{\ }$%
\boldsymbol{MT}$.

\subsection{Appendix A: $\mathbb{D}_{3}$ symmetry}

In this appendix, we review useful aspects of the symmetries of the
triangles (\ref{ta},\ref{M3}) by focussing on the example of the equilateral
triangle as it has richer symmetries. We describe the relationships between
the axial $\boldsymbol{C}_{3}^{z}$ of the equilateral triangle cross-section
of the Figures \ref{e} and its plane symmetries $\boldsymbol{M}_{1},%
\boldsymbol{M}_{2},\boldsymbol{M}_{3}$. The axial $\boldsymbol{C}_{3}^{z}$
symmetry of the equilateral triangle is intimately related with the $%
\boldsymbol{M}_{i}$\ reflections; for example it can be factorised as the
intersection of two plane symmetries $\boldsymbol{M}_{i}$ and $\boldsymbol{M}%
_{i+1}$ like%
\begin{equation}
\boldsymbol{C}_{3}^{z}=\boldsymbol{M}_{i+1}\boldsymbol{M}_{i}\qquad ,\qquad
i=1,2,3\text{ \  \ }\func{mod}3  \label{fc}
\end{equation}%
where the three $\boldsymbol{M}_{j}$'s are precisely the mirror plane
depicted by the Figure \textbf{\ref{T1}}. This factorisation feature can be
checked directly by starting from the set of corners of the triangle ordered
like $\left( A_{i-1},A_{i},A_{i+1}\right) $; and perform two successive
mirror symmetries: First; apply the plane symmetry $\boldsymbol{M}_{i}$
fixing $A_{i}$ and permuting $\left \{ A_{i-1},A_{i+1}\right \} $; this
mapping gives $\left( A_{i+1},A_{i},A_{i-1}\right) $. Then, act by $%
\boldsymbol{M}_{i+1}$ fixing $A_{i+1}$ and permuting the others namely $%
\left \{ A_{i-1},A_{i}\right \} $; this leads to $\left(
A_{i+1},A_{i-1},A_{i}\right) $ which is nothing but a cyclic $\boldsymbol{C}%
_{3}^{z}$ transformation of $\left( A_{i-1},A_{i},A_{i+1}\right) .$ However,
the factorisation (\ref{fc}) is deeper and deserves a discussion; it has an
interesting interpretation in terms of the Dihedral $\mathbb{D}_{3}$ having
six elements including the order 3 cyclic $\mathbb{Z}_{3}$ --- rotations by $%
\frac{2\pi r}{3}$ --- and the three reflections $\boldsymbol{M}_{1},%
\boldsymbol{M}_{2},\boldsymbol{M}_{3}$; i.e: 
\begin{equation}
\mathbb{D}_{3}=\left \{ \boldsymbol{I}_{id},\boldsymbol{C}_{3}^{z},(%
\boldsymbol{C}_{3}^{z})^{-1},\boldsymbol{M}_{1},\boldsymbol{M}_{2},%
\boldsymbol{M}_{3}\right \}
\end{equation}%
Besides itself and identity, this set has four other subgroups; the cyclic $%
\mathbb{Z}_{3}$ given by $\{ \boldsymbol{I}_{id},\boldsymbol{C}_{3}^{z},(%
\boldsymbol{C}_{3}^{z})^{-1}\}$ with generator $\boldsymbol{C}_{3}^{z}$, and
three $\mathbb{Z}_{2}$'s given by \{$\boldsymbol{I}_{id},\boldsymbol{M}_{i}$%
\} and generated by the reflections $\boldsymbol{M}_{i}$,%
\begin{equation}
\begin{tabular}{|l|l|l|l|l|}
\hline
subgroups & I$_{id}$ & $\mathbb{Z}_{2}$ & $\mathbb{Z}_{3}$ & $\mathbb{D}_{3}$
\\ \hline
\ number & 1 & 3 & 1 & 1 \\ \hline
\end{tabular}
\label{bt}
\end{equation}%
\begin{equation*}
\end{equation*}%
Moreover, as $\mathbb{D}_{3}$\ is isomorphic to the permutation group $%
\mathbb{S}_{3}$ of three elements; say $\left \{ a,b,c\right \} $ ---for the
three corners A$_{1},$A$_{2},$A$_{3}$ ---, one can write explicitly these
elements by using cycles. In this language, $\boldsymbol{C}_{3}^{z}$ is
viewed as a particular element of $\mathbb{S}_{3}$ amongst the $3!=6$
possible ones; they include two 3-cycles $\left( abc\right) $ and $\left(
acb\right) $; three transpositions $\left( ab\right) ,$ $\left( ac\right) ,$ 
$\left( bc\right) ;$ and the identity $\left( a\right) \left( b\right)
\left( c\right) $. The $\boldsymbol{C}_{3}^{z}$ and its inverse ($%
\boldsymbol{C}_{3}^{z})^{-1}$ are given by the 3-cycle $\left( abc\right) $
and its inverse $\left( acb\right) $; $\boldsymbol{C}_{3}^{z}$ generates the 
$\mathbb{Z}_{3}$ subgroup in above table. The three reflections $\boldsymbol{%
M}_{1},$ $\boldsymbol{M}_{2},$ $\boldsymbol{M}_{3}$ are given by the simple
transpositions $\left( ab\right) ,$ $\left( ac\right) ,$ $\left( bc\right) $%
; they generate the three $\mathbb{Z}_{2}$'s in (\ref{bt}). Recall that the
6 elements of the group $\mathbb{S}_{3}$ can be generated by two non
commuting transpositions; say $\left( cb\right) $ and $\left( ac\right) $;
we have for example 
\begin{equation}
\left( abc\right) =\left( cb\right) \left( ac\right)
\end{equation}%
which can be compared with (\ref{fc}). Notice moreover that being a 3-cycle,
the $\boldsymbol{C}_{3}^{z}$ permits in turn to relate the $\boldsymbol{M}%
_{1},$ $\boldsymbol{M}_{2},$ $\boldsymbol{M}_{3}$ planes to each other as
follows%
\begin{equation}
\boldsymbol{C}_{3}^{z}\boldsymbol{M}_{i}=\boldsymbol{M}_{i+1}\qquad
with\qquad i=1,2,3\text{ \  \ }\func{mod}3  \label{mm}
\end{equation}%
This conjugation property is interesting for our analysis; it shows that it
is enough to restrict the study to one of the three mirror symmetries; say $%
\boldsymbol{M}_{1}$; in agreement with the observation of \textrm{\cite{VOP}}%
; the results for the two others follow\ by applying the above conjugation
relation.\ 

\subsection{Appendix B: Hexagonal frame}

We begin by noticing that the computation of the components $\mathcal{F}_{a}$%
's in (\ref{hf}) is some how cumbersome when they are expressed in the cubic
frame with momentum vector $\mathbf{k}$ decomposed as $k_{x}\mathbf{e}%
_{x}+k_{y}\mathbf{e}_{y}+k_{z}\mathbf{e}_{z}$. This is because of the
triangular symmetry of the tri-hinge cylinder; so it is interesting to take
advantage of this property by using a basis vector frame exhibiting this
symmetry namely the hexagonal frame \textrm{\cite{NPB,PRD1}}. In practice,
this can be done by first splitting the momentum vector $\mathbf{k}$ ( 
\textrm{and generally speaking 3- vectors} $\mathbf{\upsilon }$), like $%
\mathbf{k}_{\Vert }+k_{z}\mathbf{e}_{z}$ ( $\mathbf{\upsilon =\upsilon }%
_{\Vert }+\upsilon _{z}\mathbf{e}_{z}$); then express the transverse $%
\mathbf{k}_{\Vert }$ in the hexagonal basis generated by two vectors planar $%
\mathbf{\alpha }_{1},\mathbf{\alpha }_{2}$ as follows $\mathbf{k}_{\Vert
}=q_{1}\mathbf{\alpha }_{1}+q_{2}\mathbf{\alpha }_{2}$. In the new basis,
the momentum components $q_{1}$ and $q_{2}$ propagate along $\mathbf{\alpha }%
_{1}$ and $\mathbf{\alpha }_{2}$ directions in similar way to $%
k_{x},k_{y},k_{z}$ which propagate in the cubic $\mathbf{e}_{x},$ $\mathbf{e}%
_{y},$ $\mathbf{e}_{z}$ directions; these q- components are obviously
functions of the old $k_{x}$ and $k_{y}$ with relationships as in \textrm{eq(%
\ref{qq})}. To engineer domain walls and exhibit the composed symmetry $%
\boldsymbol{M}\boldsymbol{T}$, we need other tools whose useful ones for our
calculus are described here after. First, recall that the hexagonal basis
has a metric $K_{ij}=\mathbf{\alpha }_{i}.\mathbf{\alpha }_{j}$ which is
different from the usual cubic metric $\delta _{ij}=\mathbf{e}_{i}.\mathbf{e}%
_{j}$; by implementing z-direction, the $K$ reads as%
\begin{equation}
K=\left( 
\begin{array}{ccc}
\frac{4}{3} & -\frac{2}{3} & 0 \\ 
-\frac{2}{3} & \frac{4}{3} & 0 \\ 
0 & 0 & 1%
\end{array}%
\right) \qquad ,\qquad K^{-1}=\left( 
\begin{array}{ccc}
1 & \frac{1}{2} & 0 \\ 
\frac{1}{2} & 1 & 0 \\ 
0 & 0 & 1%
\end{array}%
\right)
\end{equation}%
Notice also that, contrary to the cubic $\mathbf{e}_{i}$- vector basis, the
hexagonal $\mathbf{\alpha }_{i}$'s are not reflexive vectors in the sense
that one needs also their $\mathbf{\omega }_{i}$\ dual with metric $%
K_{ij}^{-1}=\mathbf{\omega }_{i}.\mathbf{\omega }_{j}$; the extra $\mathbf{%
\omega }_{i}$'s are needed to do projections by using the duality property $%
\mathbf{\alpha }_{i}.\mathbf{\omega }_{j}=\delta _{ij}$; this is not
necessary in the cubic frame as we already have $\mathbf{e}_{i}.\mathbf{e}%
_{j}=\delta _{ij}$. A realisation of these $\mathbf{\alpha }_{i}$'s and $%
\mathbf{\omega }_{i}$'s in the $\mathbf{e}_{i}$-cubic basis is given by 
\begin{equation}
\begin{tabular}{lllllll}
$\mathbf{\alpha }_{1}$ & $=$ & $(1,+\frac{1}{\sqrt{3}},0)$ & $\qquad ,\qquad 
$ & $\mathbf{\omega }_{1}$ & $=$ & $(\frac{1}{2},+\frac{\sqrt{3}}{2},0)$ \\ 
$\mathbf{\alpha }_{2}$ & $=$ & $(1,-\frac{1}{\sqrt{3}},0)$ & $\qquad ,\qquad 
$ & $\mathbf{\omega }_{2}$ & $=$ & $(\frac{1}{2},-\frac{\sqrt{3}}{2},0)$%
\end{tabular}
\label{12}
\end{equation}%
\begin{equation*}
\end{equation*}%
having a mirror symmetry; under the reflection $\left( \mathbf{e}_{x},%
\mathbf{e}_{y}\right) \rightarrow \left( \mathbf{e}_{x},-\mathbf{e}%
_{y}\right) $, the vectors of the pair $\left( \mathbf{\alpha }_{1},\mathbf{%
\alpha }_{2}\right) $ gets transformed into $\left( \mathbf{\alpha }%
_{1}^{\prime },\mathbf{\alpha }_{2}^{\prime }\right) =\left( \mathbf{\alpha }%
_{2},\mathbf{\alpha }_{1}\right) $; the same thing holds for the dual basis
as we also have $\left( \mathbf{\omega }_{1}^{\prime },\mathbf{\omega }%
_{2}^{\prime }\right) =\left( \mathbf{\omega }_{2},\mathbf{\omega }%
_{1}\right) $. Notice also that (\ref{12}) is not the unique $\mathbb{Z}_{2}$
symmetric way to expand $\mathbf{k}_{\Vert }$; there are two other
remarkable basis sets that we comment here below considering that they are
interesting in the exhibition of the three $\mathbb{Z}_{2}$ mirror
symmetries (\ref{sy}) which act by permuting the basis vector directions 
\textrm{\cite{PRD1,PRD2}}. Instead of the pair $\left( \mathbf{\alpha }_{1},%
\mathbf{\alpha }_{2}\right) $, one can also consider either the pair $\left( 
\mathbf{\alpha }_{1},\mathbf{\alpha }_{3}\right) $ or the pair $\left( 
\mathbf{\alpha }_{2},\mathbf{\alpha }_{3}\right) $; the three planar vectors
making these three pairs are linked by the relation $\mathbf{\alpha }_{1}+%
\mathbf{\alpha }_{2}+\mathbf{\alpha }_{3}=0$ which is invariant under $%
\mathbb{D}_{3}$. The same thing can be said about $\mathbf{\omega }%
_{i}^{\prime }$s with $\mathbf{\omega }_{1}+\mathbf{\omega }_{2}+\mathbf{%
\omega }_{3}=0$. \newline
By using these hexagonal vector bases, we derive useful relations for our
study; for instance by expressing momentum $\mathbf{k}$ like $q_{1}\mathbf{%
\alpha }_{1}+q_{2}\mathbf{\alpha }_{2}+k_{z}\mathbf{e}_{z}$, we discover (%
\ref{qq}) with $q_{1}+q_{2}+q_{3}=0$. This constraint captures the full $%
\mathbb{D}_{3}$ symmetry of the triangular section; and is inserted in our
calculations through the Dirac delta function $\delta _{\left( f\right) }$
where we have set $f=q_{1}+q_{2}+q_{3}$; for the simplicity of the
presentation, we shall hide this delta function. Notice also that under the
reflection $\left( k_{x},k_{y}\right) \rightarrow \left( k_{x},-k_{y}\right) 
$, the triplet $\left( q_{1},q_{2},q_{3}\right) $ gets mapped to $\left(
q_{2},q_{1},q_{3}\right) .$ An example of a function with dependence into
the three $q_{i}$'s is given by the following complex function $Z\left(
q_{1},q_{2},q_{3}\right) $ that turns out to play an important role in our
study%
\begin{equation}
Z=e^{2i\pi q_{1}}+e^{2i\pi q_{2}}+e^{2i\pi q_{3}}  \label{z}
\end{equation}%
Because of the Dirac delta function, we can present this $Z$ in three
manners; one of them is given by solving the condition $f=0$ like $%
q_{3}=-q_{1}-q_{2}$ and put it back into (\ref{z}); another manner is by
substituting $q_{2}=-q_{1}-q_{3}$. The real part of above complex function Z
gives precisely the $F_{x}$- component involved in the hamiltonian (\ref{hf}%
). To complete this digression on the tools and the symmetry properties of
the F$_{a}$'s, notice also that there are two more decompositions: one
concerns the gamma matrix vector $\mathbf{\vec{\gamma}}$ and the other
regards the way to compute the scalar product $\vec{F}.\mathbf{\vec{\gamma}}$
appearing in (\ref{hf}). In the cubic frame, $\mathbf{\vec{\gamma}}$
decomposes as usual like $\mathbf{\gamma }_{x}\mathbf{\vec{e}}_{x}+\mathbf{%
\gamma }_{y}\mathbf{\vec{e}}_{y}+\mathbf{\gamma }_{z}\mathbf{\vec{e}}_{z}$
and so the scalar $\vec{F}.\mathbf{\vec{\gamma}}$ can be expanded as $F_{x}%
\mathbf{\gamma }_{x}+F_{y}\mathbf{\gamma }_{y}+F_{z}\mathbf{\gamma }_{z}$.
Here also it is useful to decompose the vectors $\vec{F}$ and $\mathbf{\vec{%
\gamma}}$ in the hexagonal bases; first, we split the vector operator $%
\mathbf{\vec{\gamma}}$ like the sum $\mathbf{\vec{\gamma}}_{\Vert }+\mathbf{%
\gamma }_{z}\mathbf{\vec{e}}_{z}$ and then decompose the transverse $\mathbf{%
\vec{\gamma}}_{\Vert }$ as follows $\Gamma _{1}\mathbf{\vec{\omega}}%
_{1}+\Gamma _{2}\mathbf{\vec{\omega}}_{2}$ where the new matrices $\Gamma
_{1}\mathbf{,}\Gamma _{2}$ are linear combinations of the old gamma matrices 
$\mathbf{\gamma }_{x}\mathbf{,\gamma }_{y}$. Regarding $\vec{F}$, it is
convenient to decompose in same manner as we have done for momentum vector
namely like $G_{1}\mathbf{\vec{\alpha}}_{1}+G_{2}\mathbf{\vec{\alpha}}%
_{2}+k_{z}\mathbf{\vec{e}}_{z}$ where the new $G_{1}\mathbf{,}G_{2}$ are
functions of $F_{x}\mathbf{,}F_{y}$. To compute the various coefficients, we
use the projections $G_{i}=\vec{F}.\mathbf{\vec{\omega}}_{i}$ and $\Gamma
_{i}=\mathbf{\vec{\gamma}}.\mathbf{\vec{\alpha}}_{i}$, that when put back
into (\ref{hf}), we obtain 
\begin{equation}
H=G_{1}\mathbf{\Gamma }_{1}+G_{2}\mathbf{\Gamma }_{2}+\Delta _{z}\left( \sin
k_{z}\right) \mathbf{\gamma }_{z}+F_{4}\mathbf{\gamma }_{4}+F_{5}\mathbf{%
\gamma }_{5}  \label{fg}
\end{equation}%
The block $G_{1}\mathbf{\Gamma }_{1}+G_{2}\mathbf{\Gamma }_{2}$ is obviously
equal to $F_{x}\mathbf{\gamma }_{x}+F_{x}\mathbf{\gamma }_{y}$ as the scalar
product of vectors is frame independent. \newline
In the end of this appendix, notice that the $\varrho _{i}$ projectors used
in section 5 are given by $\left \vert i\right \rangle \left \langle
i\right
\vert $; their matrix forms are 
\begin{equation}
\varrho _{1}=\left( 
\begin{array}{ccc}
1 & 0 & 0 \\ 
0 & 0 & 0 \\ 
0 & 0 & 0%
\end{array}%
\right) \quad ,\quad \varrho _{2}=\left( 
\begin{array}{ccc}
0 & 0 & 0 \\ 
0 & 1 & 0 \\ 
0 & 0 & 0%
\end{array}%
\right) \quad ,\quad \varrho _{3}=\left( 
\begin{array}{ccc}
0 & 0 & 0 \\ 
0 & 0 & 0 \\ 
0 & 0 & 1%
\end{array}%
\right)
\end{equation}%
and they transform under the $\mathbb{D}_{3}$ transformations as collected
in the following table%
\begin{equation}
\begin{tabular}{|l|l|l|l|l|}
\hline
& $\varrho _{1}$ & $\varrho _{2}$ & $\varrho _{3}$ & $\varrho _{0}$ \\ \hline
$\boldsymbol{M}_{1}=\boldsymbol{t}_{12}=\left( 12\right) $ & $\varrho _{2}$
& $\varrho _{1}$ & $\varrho _{3}$ & $\varrho _{0}$ \\ \hline
$\boldsymbol{M}_{2}=\boldsymbol{t}_{23}=\left( 23\right) $ & $\varrho _{1}$
& $\varrho _{3}$ & $\varrho _{2}$ & $\varrho _{0}$ \\ \hline
$\boldsymbol{M}_{3}=\boldsymbol{t}_{31}=\left( 31\right) $ & $\varrho _{3}$
& $\varrho _{2}$ & $\varrho _{1}$ & $\varrho _{0}$ \\ \hline
$\boldsymbol{C}_{3}^{z}=\left( 123\right) $ & $\varrho _{2}$ & $\varrho _{3}$
& $\varrho _{1}$ & $\varrho _{0}$ \\ \hline
\end{tabular}
\label{rr}
\end{equation}

\subsection{Appendix C: Deformations preserving\emph{\ }$\boldsymbol{MT}$}

In this appendix, we start from the eight band model hamiltonian $H=H_{%
\mathbf{k}}$ given by Eq(\ref{hf}) preserving\emph{\ }$\boldsymbol{MT}$; and
we make comments regarding its deformations $\delta H_{\mathbf{k}}$ due to
external fields. We show that there are \textrm{59} operators generating $%
\delta H_{\mathbf{k}}$, part of them preserves $\boldsymbol{MT}$ and the
other part breaks $\boldsymbol{MT}$; their number depend on intrinsic data
of $\delta H_{\mathbf{k}}$. Before going into details, we would like to
notice that a complete theoretical description of the deformations of $H_{%
\mathbf{k}}$ requires involved tools on the algebra of 8$\times $8 Dirac
gamma matrices as they encode the information the $\delta H_{\mathbf{k}}$
deviations of $H_{\mathbf{k}}$; they need also to specify the magnitudes of
the coupling parameters in the tight binding model (TBM) relying on the
stacked lattices depicted by the Figure \textbf{\ref{ab}}. Below, we will
avoid complexity and focus mainly on describing the method and commenting
results. To that purpose, we first describe some aspects of $H_{\mathbf{k}}$%
; then we turn to study the deviations $\delta H_{\mathbf{k}}$, due to
perturbations, and the first step towards their full classifications
according to their charges under $\boldsymbol{MT};$ i.e whether $H_{\mathbf{k%
}}$ preserves $\boldsymbol{MT}$ or not. For later use, we also recall that
the TRS generator $\boldsymbol{T}$ is given by $\zeta _{0}\tau _{0}\sigma
_{y}\boldsymbol{K}$ while the mirror $\boldsymbol{M}$ has been realised as $%
\zeta _{0}\tau _{0}\sigma _{z}$; so $\boldsymbol{MT}=i\zeta _{0}\tau
_{0}\sigma _{x}\boldsymbol{K}$.

$\bullet $ \emph{Some useful properties of} $H_{\mathbf{k}}$\newline
We start by recalling that the $H_{\mathbf{k}}$ of Eq(\ref{hf})\ is a
hermitian 8$\times $8 matrix having some special properties that are useful
for the study of its deformations $\delta H_{\mathbf{k}}$ as well as their
protection symmetries. We describe in what follows \textrm{four} of these
properties while aiming for $\delta H_{\mathbf{k}}$ and its invariance: $%
\left( 1\right) $ The $H_{\mathbf{k}}$ we considered involves 5 terms which
may be imagined as given by the following formal sum%
\begin{equation}
H_{\mathbf{k}}=\mathcal{F}_{1}\boldsymbol{X}_{1}+\mathcal{F}_{2}\boldsymbol{X%
}_{2}+\mathcal{F}_{3}\boldsymbol{X}_{3}+\mathcal{F}_{4}\boldsymbol{X}_{4}+%
\mathcal{F}_{5}\boldsymbol{X}_{5}  \label{hk}
\end{equation}%
where the $\mathcal{F}_{i}^{\prime }=\mathcal{F}_{i}\left( \mathbf{k}%
,\Lambda \right) $ are real functions of momentum $\mathbf{k}$ and the
hopping parameters $\Lambda $ as in Eq(\ref{hf}). $\left( 2\right) $ The $%
\boldsymbol{X}_{i}$'s are five hermitian 8$\times $8 matrices which are
realised in terms of tensor products of three sets of Pauli matrices as $%
\zeta _{\alpha }\tau _{a}\sigma _{i}$; these products are collected in the
following table\footnote{%
\ Along with these 5 matrices $\boldsymbol{X}_{1},\boldsymbol{X...,X}_{5}$,
we have moreover two other particular matrices: a sixth $\boldsymbol{X}%
_{6}=\zeta _{x}\tau _{z}\sigma _{0}$ and a seventh $\boldsymbol{X}_{7}=\zeta
_{y}\tau _{z}\sigma _{0}$; these matrices haven't been used in Eq(\ref{hf}).
Notice also that we have $\boldsymbol{X}_{7}=i\boldsymbol{X}_{1}\boldsymbol{X%
}_{2}\boldsymbol{X}_{3}\boldsymbol{X}_{4}\boldsymbol{X}_{5}\boldsymbol{X}%
_{6}.$}%
\begin{equation}
\begin{tabular}{|l|l|l|l|l|l|}
\hline
\ {\small generators} & $\  \  \boldsymbol{X}_{1}$ & $\  \boldsymbol{X}_{2}$ & $%
\  \boldsymbol{X}_{3}$ & $\  \boldsymbol{X}_{4}$ & $\boldsymbol{X}_{5}$ \\ 
\hline
{\small a realisation} & $\zeta _{x}\tau _{z}\sigma _{0}$ & $\zeta _{0}\tau
_{x}\sigma _{y}$ & $\zeta _{0}\tau _{x}\sigma _{z}$ & $\zeta _{0}\tau
_{y}\sigma _{0}$ & $\zeta _{z}\tau _{z}\sigma _{0}$ \\ \hline
\end{tabular}
\label{tx}
\end{equation}%
from which we learn that $\boldsymbol{X}_{1},\boldsymbol{X}_{3},\boldsymbol{X%
}_{5}$ are real (even under $\boldsymbol{K}$) whilst $\boldsymbol{X}_{2},%
\boldsymbol{X}_{4}$ are imaginary (odd under $\boldsymbol{K}$). Notice that
the table (\ref{tx}) is not complete as\ there are 59 other 8$\times $8
matrix generators which do not figure in it; and which could be added to Eq(%
\ref{hk}); they will described later on as they are precisely the generators
of $\delta H_{\mathbf{k}}$. $\left( 3\right) $ Generally speaking, we do not
need an explicit realisation of the $\boldsymbol{X}_{i}$'s as in above
table; all we need to know is their algebraic properties which are given by: 
$\left( i\right) $ the Clifford algebra reading as $\boldsymbol{X}_{A}%
\boldsymbol{X}_{B}+\boldsymbol{X}_{B}\boldsymbol{X}_{A}=2\delta _{AB}$ with $%
A,B=1,...,6$; $\left( ii\right) $ the properties of the $\boldsymbol{X}_{A}$%
's under complex conjugation operator $\boldsymbol{K}$ as it appears in TRS
transformations; and $\left( iii\right) $ for chiral models, we also need
their properties under the operator $\boldsymbol{X}_{7}=i\boldsymbol{X}_{1}%
\boldsymbol{X}_{2}\boldsymbol{X}_{3}\boldsymbol{X}_{4}\boldsymbol{X}_{5}%
\boldsymbol{X}_{6}$. These aspects can be learnt form the domain walls
prescription of our tri-hinge system requiring the following charges under
TRS, mirror $\boldsymbol{M}$ and their composition $\boldsymbol{MT}$ 
\begin{equation}
\begin{tabular}{|l|l|l|l|l|l|}
\hline
\ {\small factors} & $\mathcal{F}_{1}\boldsymbol{X}_{1}$ & $\mathcal{F}_{2}%
\boldsymbol{X}_{2}$ & $\mathcal{F}_{3}\boldsymbol{X}_{3}$ & $\mathcal{F}_{4}%
\boldsymbol{X}_{4}$ & $\mathcal{F}_{5}\boldsymbol{X}_{5}$ \\ \hline
$\  \  \  \boldsymbol{T}$ & $\  \ +$ & $\  \  \ +$ & $\  \  \ +$ & $\  \  \ -$ & $\  \
+ $ \\ \hline
$\  \  \  \boldsymbol{M}$ & $\  \ +$ & $\  \ +$ & $\  \ +$ & $\  \  \ -$ & $\  \ +$
\\ \hline
$\  \  \boldsymbol{MT}$ & $\  \ +$ & $\  \ +$ & $\  \ +$ & $\  \  \ +$ & $\  \ +$ \\ 
\hline
\end{tabular}
\label{tm}
\end{equation}%
From this domain walls prescription as well as the trigonal symmetries of
the materials discussed in the heart of the paper requiring particular T-
and M- charges for the $\mathcal{F}_{A}$ coefficients under TRS and mirror $%
\boldsymbol{M}$, we can re-work out the realisation of $\boldsymbol{T}$ and $%
\boldsymbol{M}$. For later use, we denote the T-charge of the pair $\left( 
\mathcal{F}_{i},\boldsymbol{X}_{i}\right) $ under TRS as $\left(
q_{i},p_{i}\right) _{T}$ so that the T- charge of the product $\mathcal{F}%
_{i}\boldsymbol{X}_{i}$ is given by $q_{iT}\times p_{iT}$. Similarly, we
denote by $\left( q_{i},p_{i}\right) _{M}$ the M- charge of the pair $\left( 
\mathcal{F}_{i},\boldsymbol{X}_{i}\right) $ under $\boldsymbol{M}$ so that
the M-charge of the monomials $\mathcal{F}_{i}\boldsymbol{X}_{i}$ is given
by $q_{iM}\times p_{iM}.$ Therefore, the charge of $\mathcal{F}_{i}%
\boldsymbol{X}_{i}$ under the composite $\boldsymbol{MT}$ is given by $%
q_{iT}\times q_{iM}\times p_{iT}\times p_{iM}$. These T- and M- charges are
collected in the following table%
\begin{equation}
\begin{tabular}{|l|l|l|l|l|l|}
\hline
\ {\small factors} & ${\small (}\mathcal{F}_{1},\boldsymbol{X}_{1}{\small )}$
& ${\small (}\mathcal{F}_{2},\boldsymbol{X}_{2}{\small )}$ & ${\small (}%
\mathcal{F}_{3},\boldsymbol{X}_{3}{\small )}$ & ${\small (}\mathcal{F}_{4},%
\boldsymbol{X}_{4}{\small )}$ & ${\small (}\mathcal{F}_{5},\boldsymbol{X}_{5}%
{\small )}$ \\ \hline
{\small T- charge} & $(+,+)_{T}$ & $(-,-)_{T}$ & $(-,-)_{T}$ & $(+,-)_{T}$ & 
$(+,+)_{T}$ \\ \hline
{\small M- charge} & $(+,+)_{M}$ & $(-,-)_{M}$ & $(+,+)_{M}$ & $(-,+)_{M}$ & 
$(+,+)_{M}$ \\ \hline
{\small M- charge} & $(+,+)_{MT}$ & $(+,+)_{MT}$ & $(-,-)_{MT}$ & $%
(-,-)_{MT} $ & $(+,+)_{MT}$ \\ \hline
\end{tabular}
\label{cn}
\end{equation}%
from which we learn%
\begin{equation}
\boldsymbol{T=-i\boldsymbol{X}_{3}\boldsymbol{X}_{6}K\qquad },\boldsymbol{%
\qquad M=i\boldsymbol{X}_{2}X}_{6}\boldsymbol{\qquad },\boldsymbol{\qquad MT=%
\boldsymbol{X}_{3}\boldsymbol{X}_{2}K}
\end{equation}%
By substituting the realisation (\ref{tx}) back into above quantities, we
deduce the expressions of the symmetry generators of section 4 namely $%
\boldsymbol{T}=\zeta _{0}\tau _{0}\sigma _{y}\boldsymbol{K}$ and $%
\boldsymbol{M}=\zeta _{0}\tau _{0}\sigma _{z}$ as well as $\boldsymbol{MT}%
=i\zeta _{0}\tau _{0}\sigma _{x}\boldsymbol{K}$. $\left( 4\right) $ As
noticed earlier, the five $\boldsymbol{X}_{i}$'s obey the Clifford algebra $%
\boldsymbol{X}_{i}\boldsymbol{X}_{j}+\boldsymbol{X}_{i}\boldsymbol{X}%
_{i}=2\delta _{ij}$; this feature teaches us that $H_{\mathbf{k}}^{2}=E_{%
\mathbf{k}}^{2}I_{8}$ showing that the square of the hamiltonian is
proportional to the 8$\times $8 identity matrix $I_{8}$. This property
indicates that the eight energy eigenvalues $E_{1},...,E_{8}$ of the
hamiltonian are four fold degenerate; i.e two different energies $E_{\pm }$;
each with multiplicity 4; these energies are given by $E_{\pm }=\pm \frac{1}{%
2}E_{g}$ with gap energy $E_{g}=2\sqrt{\mathcal{F}_{1}^{2}+...+\mathcal{F}%
_{5}^{2}}$. The vanishing properties of this gap has been studied and
commented in the core of the paper.

$\bullet $ \emph{Building} $\delta H_{\mathbf{k}}$ \emph{and describing its
properties}\newline
We begin by noticing that the hamiltonian (\ref{hk}) modeling the
topological helical states is clearly not the general hamiltonian we may
write down; this is because Eq(\ref{hk}) has only five degrees of freedom $%
\mathcal{F}_{1},\mathcal{F}_{2},\mathcal{F}_{3},\mathcal{F}_{4},\mathcal{F}%
_{5}$; while the most general form of a 8$\times $8 hermitian matrix has $64$
components as exhibited by the following expansion 
\begin{equation}
\begin{tabular}{lll}
$\tilde{H}_{\mathbf{k}}$ & $=$ & $\mathcal{F}^{0}\boldsymbol{X}_{0}+\sum
\limits_{A=1}^{6}\left( \mathcal{F}^{A}\boldsymbol{X}_{A}+i\mathcal{\tilde{F}%
}^{A}\boldsymbol{X}_{A}\boldsymbol{X}_{7}\right) +\sum \limits_{A<B<C=1}^{6}%
\mathcal{G}^{AB}\boldsymbol{X}_{ABC}$ \\ 
&  & $+\sum \limits_{A<B=1}^{6}\left( \mathcal{F}^{AB}\boldsymbol{X}_{AB}+%
\mathcal{\tilde{F}}^{AB}\boldsymbol{X}_{AB}\boldsymbol{X}_{7}\right) +%
\mathcal{\tilde{F}}^{0}\boldsymbol{X}_{7}$%
\end{tabular}
\label{ex1}
\end{equation}%
where $\boldsymbol{X}_{0}=I_{8}$ and where we have set $\boldsymbol{X}_{AB}=i%
\boldsymbol{X}_{A}\boldsymbol{X}_{B}$ and $\boldsymbol{X}_{ABC}=i\boldsymbol{%
X}_{A}\boldsymbol{X}_{B}\boldsymbol{X}_{C}$; this basis is a standard basis
useful in dealing with Clifford algebra and fermionic wave functions. This
decomposition of $\tilde{H}_{\mathbf{k}}$ corresponds to the following
partition of the number 64,%
\begin{equation}
\begin{tabular}{|l|l|l|l|l|l|l|l|}
\hline
generators & $I_{8}$ & $\boldsymbol{X}_{A}$ & $\boldsymbol{X}_{AB}$ & $%
\boldsymbol{X}_{ABC}$ & $\boldsymbol{X}_{AB}\boldsymbol{X}_{7}$ & $%
\boldsymbol{X}_{A}\boldsymbol{X}_{7}$ & $\boldsymbol{X}_{7}$ \\ \hline
$\  \  \  \ 64$ & $1$ & $\ 6$ & $15$ & $\ 20$ & $\  \  \  \ 15$ & $\  \  \  \ 6$ & $1$
\\ \hline
\end{tabular}
\label{cc}
\end{equation}%
Contrary to Eq(\ref{hk}), the hamiltonian $\tilde{H}_{\mathbf{k}}$ has in
general eight different energy eigenvalues that can be presented like $%
E_{1}^{\pm },E_{2}^{\pm },E_{3}^{\pm },E_{4}^{\pm }$; the four $E_{i}^{+}$
stand for the energies of conducting bands while the four $E_{i}^{-}$ give
the energies of valence bands; the gap energy $E_{g}$ is given by $E_{\min
}^{+}-E_{\max }^{-}$; and then the knowledge of these $E_{\min }^{+}$ and $%
E_{\max }^{-}$ is of major importance for several issues; in particular the
one regarding those small perturbations breaking $\boldsymbol{MT}$ and their
effect on the gap energy. Here, we will not calculate $E_{\min }^{+}$ nor $%
E_{\max }^{-}$ as they require specifying magnitudes of the coupling
constants in TBM; and also solving a somehow complicated eigenvalue problem
which, though interesting numerically, is beyond the main objective of the
paper. \newline
In summary, we end up this appendix by saying that the hamiltonian Eq(\ref%
{hk}) modeling helical gapless states in tri-hinge system is a candidate
hamiltonian having three remarkable properties: $\left( i\right) $ $%
\boldsymbol{MT}$ protection; $\left( ii\right) $ admitting two energy
eigenvalues with multiplicity 4; and $\left( iii\right) $ a gap energy $%
E_{g}=E_{\min }^{+}-E_{\max }^{-}$ which can take a zero value at Dirac
points and for some values of the coupling parameters. The general $\tilde{H}%
_{\mathbf{k}}$ given by (\ref{ex1}) goes beyond these properties and may be
thought of as $H_{\mathbf{k}}+\delta H_{\mathbf{k}}$; that is describing the
deviations of $H_{\mathbf{k}}$. So, the $\delta H_{\mathbf{k}}$ is generated
by 59 monomial operators as listed by (\ref{cc}). Moreover, as for the T-
and M- charges (\ref{cn}), one can work out a classification of the
deformations $\delta H_{\mathbf{k}}$; they depend on the charges of the
coefficients $\mathcal{F}^{A_{1}...A_{i}}\left( \mathbf{k},\Lambda
,..\right) $ in the expansion (\ref{ex1}). This is a technical question
which requires a bigger space for an appendix; we hope to return to it in a
future occasion. We conclude by noting that the deformations $\delta H_{%
\mathbf{k}}$ generally lift the energy degeneracy of $H_{\mathbf{k}}$ and
the knowledge of the explicit expression of $E_{\min }^{+}-E_{\max }^{-}$
allows to study the properties of the energy gap induced by small
perturbations breaking $\boldsymbol{MT}$. 
\begin{equation*}
\end{equation*}

\begin{acknowledgement}
Professors Lalla Btissam Drissi and El Hassan Saidi would like to
acknowledge \textquotedblright Acad\'{e}mie Hassan II des Sciences et
Techniques-Morocco\textquotedblright \ for financial support. They also
thank Felix von Oppen for stimulating discussions. L. B. Drissi acknowledges
the Alexander von Humboldt Foundation for financial support via the Georg
Forster Research Fellowship for experienced scientists (Ref 3.4 - MAR -
1202992).
\end{acknowledgement}

\end{document}